\documentstyle[aps,prd,eqsecnum,epsf]{revtex}
\title{
Gravitational waves from a test particle \break
scattered by a neutron star: Axial mode case
}

\author{
Kazuhiro Tominaga$^{1}$
\thanks{Electronic address:tominaga@gravity.phys.waseda.ac.jp}, 
Motoyuki Saijo$^{1}$
\thanks{Electronic address:saijo@gravity.phys.waseda.ac.jp}, 
and
Kei-ichi Maeda$^{1,2}$
\thanks{Electronic address:maeda@gravity.phys.waseda.ac.jp}
}

\address{
$^{1}$ Department of Physics,  
Waseda University, 
3-4-1 Okubo, Shinjuku, Tokyo 169-8555, Japan
}

\address{
$^{2}$ Advanced Research Institute for Science and Engineering, \\
Waseda University, Shinjuku, Tokyo 169-8555, Japan
}

\date{\today}

\begin{document}

\maketitle

%%%%%%%%%%%%%%%%%%%%%%%%%%%%%%%%%%%%%%%%%%%%%%%%%
\begin{abstract}
%%%%%%%%%%%%%%%%%%%%%%%%%%%%%%%%%%%%%%%%%%%%%%%%%
%\baselineskip 12pt 

Using a metric perturbation method, 
we study gravitational waves from a test particle 
scattered by a spherically symmetric relativistic  star.
We calculate the energy spectrum and the waveform of gravitational waves 
for axial modes.
Since metric perturbations in  axial modes do not couple to the matter
fluid of the star, emitted waves for a normal neutron star show 
only one peak in the spectrum, which  
corresponds to the orbital frequency at the turning point, where 
the gravitational field is strongest.
However, for an ultracompact star (the radius $R \lesssim 3M$),  another type of
resonant periodic peak appears
in the spectrum. This is just because of an excitation  by a scattered
particle of axial quasinormal modes, which were found  by 
Chandrasekhar and Ferrari. This
excitation comes from the existence of the potential minimum inside of a
star. We also find for an ultracompact star 
many small periodic peaks at the frequency region beyond the maximum of the potential, 
which  would be due to a resonance of
two waves reflected by two potential barriers (Regge-Wheeler type and 
one at the center of the star).
Such resonant peaks appear neither for a normal neutron star nor for a 
Schwarzschild black hole.
Consequently, even if  we analyze the energy spectrum 
of gravitational waves only for axial modes, 
it would be possible to distinguish 
between an ultracompact star and a normal neutron star (or 
a Schwarzschild black hole). \\
PACS number(s): 04.25.Nx, 04.30.-w, 04.40.Dg
\end{abstract}

%%%%%%%%%%%%%%%%%%%%%%%%%%%%%%%%%%%%%%%%%%%%%%%%%
\section{Introduction}
%%%%%%%%%%%%%%%%%%%%%%%%%%%%%%%%%%%%%%%%%%%%%%%%%

A laser interferometer to detect gravitational waves, 
such as the Laser Interferometric Gravitational Wave Observatory(LIGO), 
VIRGO, GEO600, or TAMA300,
will be in operation within several years\cite{GO}. It is one of  the
most urgent subjects for theoretical relativists to  make  a set of 
templates of gravitational waves. The direct detection of gravitational 
waves is very  important  not only as a new ``eye" observing the Universe
(gravitational wave astronomy) but also as a new probe for fundamental
physics. For example, we know that a neutron
star, which is observed as a pulsar, has a
mass $M \approx 1.4 M_{\odot}$ 
and a radius $R \approx 10$  ${\rm km}$. 
Although we can guess its central region with some theoretical ansatz, 
we have no direct information from the inside of a neutron star because
neither  radiation nor a neutrino is transparent in the central region.
The direct detection of gravitational waves may provide some information
about the inside of a neutron star when it is formed, resulting in new
observational facts or constraints about  the  equation of state for high
density matter.

One of the most promising astronomical sources for these 
ground-based detectors 
is a coalesced compact binary such as a neutron star binary.
To complete a set of 
templates from such a  source,
many studies about the emitted gravitational waves  have been done
 using various techniques.

A numerical simulation without approximations (numerical relativity) 
is probably 
one of the best ways
to calculate gravitational waves emitted 
from the compact binary system \cite{AnninosEqM,AnninosNotEqM}.
Although steady progress in numerical relativity has been
achieved, there are still some difficulties which prevent  us from 
finding the
final results.

Instead,  there are some powerful approximation methods to mimic the 
coalescence of a compact binary system.
One such approximation is the perturbative approach.
Regge and Wheeler \cite{RW} and Zerilli \cite{Zerilli}  first formulated a method for  metric
perturbations in a Schwarzschild black hole spacetime.
Then there are many works using such a black hole perturbation method
in which we  describe gravitational waves by metric perturbations 
in a black hole
background spacetime and treat a companion of the binary 
%is treated   
as a test particle\cite{Davis,OoharaFall,ON}.
 This approximation for gravitational waves in a head-on
collision of two black holes gives a good agreement 
with the results obtained by numerical simulation\cite{Anninos}. 
Thus, such a perturbative  approach may provide a good approximation for 
 gravitational waves from  a compact binary as well.

However, there are so far few works for neutron star 
perturbations. One may think that if we have two $1.4 M_{\odot}$
neutron stars, it will be a black hole after coalescence, when 
the ringing tail of the gravitational waves
can be described by 
 quasinormal modes of the black hole.
The quasinormal modes of a neutron star may have nothing to do with 
the final stage of the coalescence.
However, before black hole formation, 
we have to analyze the orbital evolution of the neutron star binary. 
Therefore,
we believe that a perturbation analysis for a neutron star is also
important.

For a nonradial pulsation of a spherically symmetric relativistic star, 
perturbation equations both for axial and polar modes
were first derived by Thorne and Campolattaro \cite{TC}.
Similar to the black hole perturbation, 
the axial mode for a spherical star is  described 
by a second order differential equation.
On the other hand, the polar mode was first described by a
fifth order differential equation\cite{TC}.
However, because of  the dynamical degrees of freedom of polar modes, 
i.e.,  sound waves and gravitational waves, the basic equation can be
reduced to a fourth order differential equation as shown by  
Lindblom and Detweiler \cite{LD}.
 Lindblom, Mendell, and Ipser \cite{LMI} also derived a couple 
of second order differential equations, which could be interpreted as 
those   for sound waves and gravitational waves, respectively.

As in the case of black hole perturbations, quasinormal modes, if  they
exist, will be important in the analysis of gravitational waves from a
compact binary.  
For the polar mode, there are two types of quasinormal modes
of a neutron  star;  one is a fluid oscillation mode similar to the
$f$,
$p$, $g$ modes of a Newtonian star (cf. \cite{Balbinski}); 
the other is a wave mode ($w$ mode),
which exists only in general relativity
\cite{Kojima,KS,LNS,AKS,AKK}. Even for the axial mode, 
Chandrasekhar and Ferrari\cite{CF} illustrated that 
 quasinormal modes  exist 
if the effective potential has a minimal  as that in an ultracompact star.
Kokkotas \cite{Kokkotas} numerically calculated such axial quasinormal modes 
for various ultracompact stars.
Kojima \cite{Kojima_sr}
extended it to a slowly rotating star, 
setting Regge-Wheeler gauge conditions.

As for emitted gravitational waves from a binary system,  
Kojima \cite{Kojima_circle}  studied  them for the case that a test
particle is moving in a circular orbit  around a polytropic star.
For a circular orbit, 
the axial mode cancels from the orbital symmetry 
so that the polar mode was only taken into account.
He showed that the resonance mode appears in the energy flux
when the orbit reaches 
the radius where the orbital frequency  
 coincides with that of the quasinormal mode.
For the axial mode, recently, Borrelli \cite{Borrelli} calculated 
 energy spectra of 
gravitational waves emitted by a test particle 
spiraling into an ultracompact 
star.
In his result, its energy spectrum shows many peaks, which correspond 
to axial quasinormal modes of the star.
Although it is important, since we are interested in a compact binary,
a direct collision might show some different features from the binary 
case. We also do not know what will happen when a test particle reaches
the surface of the star.
To avoid such an unknown factor and to analyze the case of a compact binary,
we first  study gravitational waves emitted 
when a test particle is scattered by a neutron star
in this paper.

For a Schwarzschild  black hole, Oohara and Nakamura \cite{ON} 
analyzed it and showed that  the energy spectrum has
 no peak except for one which corresponds to the orbital frequency. 
Here we consider a spherical star  instead of a Schwarzschild black hole.
We study only the axial mode 
which does not exist in Newton gravity.
To discuss the dependence of the equations of state, we analyze two models:
a uniform density star 
as in \cite{CF,Kokkotas,Borrelli} and
a polytropic star.

This paper is organized as follows. 
In Secs. \ref{sec:Linear}A and \ref{sec:Linear}B, 
we briefly review the perturbation theory for a spherically symmetric
relativistic star.  We show our numerical results in Sec. \ref{sec:GW}.
Section \ref{sec:Discussion} is devoted to a discussion. 
Some numerical techniques for 
calculating gravitational waves are summarized in the Appendix. 
Throughout this paper, we use units of $c = G = 1$ and a metric
signature  of $(-, +, +, +)$.

%%%%%%%%%%%%%%%%%%%%%%%%%%%%%%%%%%%%%%%%%%%%%%%%%
\section{Linearized Einstein equations: Axial Modes}
\label{sec:Linear}
%%%%%%%%%%%%%%%%%%%%%%%%%%%%%%%%%%%%%%%%%%%%%%%%%

We start with summarizing briefly the metric perturbation 
of a spherically symmetric relativistic  star.
A spherically symmetric  background metric $g_{\mu\nu}^{(0)} $
is described as
%%%%%%%%%
\begin{equation}
ds^{2} = g_{\mu\nu}^{(0)} dx^\mu dx^\nu =
-e^{\nu (r)} dt^2 + e^{\lambda (r)} dr^2 
+ r^2 \left( d \theta^2 + \sin^2 \theta d \phi^2 \right),
\label{eq:bgmetric}
\end{equation}
\begin{eqnarray}
e^{- \lambda(r)} = 1 - \frac{2 M(r)}{r},
\end{eqnarray}
%%%%%%%%%
where $M(r)$ is a mass function inside a radius $r$. 
We assume  a perfect fluid as the stellar matter, i.e.
%%%%%%%%%
\begin{equation}
T^{\mu\nu} = ( \rho + P ) u^{\mu} u^{\nu} + P g^{\mu \nu},
\end{equation}
%%%%%%%%%
where $\rho$ and $P$ are the total energy density and 
the pressure, respectively, which satisfy a
 barotropic equation of state, $P=P(\rho)$.
Using Tolman-Oppenheimer-Volkov equations, we construct a stellar
model and its spacetime metric numerically except for the 
constant density star case.
The radius $R$ of a star is defined by a surface 
where the pressure vanishes,
and the  mass of the star $M$ is given by $M=M(R)$.

For linearized Einstein equations,
we can  decompose the perturbed metric 
into an axial mode $h_{\mu\nu}^{\rm (axial)}$ and 
a polar mode $h_{\mu\nu}^{\rm (polar)}$:
%%%%%%%%%
\begin{equation}
g_{\mu\nu} = g_{\mu\nu}^{(0)} 
+ h_{\mu\nu}^{\rm (axial)} + h_{\mu\nu}^{\rm (polar)}.
\end{equation}
%%%%%%%%%

Using Regge-Wheeler gauge conditions, we will integrate  the interior 
and exterior regions of the
linearized Einstein equations, separately, and impose appropriate 
junction conditions at the surface $R$ to find the solution. In this
paper, we only deal with axial mode perturbation. Then, we write
down the linearized Einstein equation only for axial modes in the 
following subsections.

%%%%%%%%%%%%%%%%%%%%%%%%
\subsection{Interior region of the star}
\label{subsec:interior}
%%%%%%%%%%%%%%%%%%%%%%%%

In order to find the linearized Einstein equations for axial modes,
we expand the perturbed metric by tensor harmonics for 
polar angle $\theta$ and azimuthal angle $\phi$ \cite{Zerilli}, 
and proceed to a Fourier transformation for the time coordinate $t$.
Although  there are
ten independent components for the linearized metric,
four of them are gauge freedoms 
of coordinate transformations, which are fixed by a choice of 
 Regge-Wheeler gauge \cite{RW}, resulting in
two components belonging to axial modes, 
while the rest of the four correspond to polar modes.
For axial mode perturbations, we can set 
%%%%%%%%%
\begin{eqnarray}
h_{\mu \nu}^{(\rm axial)} dx^{\mu} dx^{\nu} 
& = & \frac{1}{2 \pi} \int d\omega \sum_{l,m} 2 h_{0,l}(r) 
\left( - \frac{1}{\sin \theta} {\partial Y_{lm}\over \partial \phi} 
dt d \theta
+ \sin \theta {\partial Y_{lm}\over \partial \theta} dt d \phi \right) 
e^{-i \omega t} \nonumber \\
&& + \frac{1}{2 \pi} \int d\omega \sum_{l,m} 2 h_{1,l}(r) 
\left( - \frac{1}{\sin \theta} {\partial Y_{lm}\over \partial \phi} 
dr d \theta
+ \sin \theta {\partial Y_{lm}\over \partial \theta} dr d \phi  \right) 
e^{-i \omega t},
\label{metric}
\end{eqnarray}
%%%%%%%%%
where $h_{0,l}$ and $h_{1,l}$ are functions 
of the radial coordinate $r$ defined below, 
and $Y_{lm}(\theta, \phi)$ is a spherical harmonics.

The  axial mode in  
the interior region of a star is described 
by a single wave equation for gravitational waves,
because a rotation does not exist in a spherically symmetric background
spacetime. That is, we have   a single wave function 
$X_{l\omega}^{\rm (int)}$, which determines the behaviors of axial 
modes as
%%%%%%%%%
\begin{eqnarray}
h_{1,l \omega} &=& e^{- (\nu - \lambda)/2} r X_{l\omega}^{\rm (int)}, \\
h_{0,l \omega} &=& \frac{i}{\omega} e^{(\nu - \lambda)/2} 
\frac{d}{dr} \left( r X_{l\omega}^{\rm (int)} \right).
\end{eqnarray}
%%%%%%%%%
The equation for the wave function 
$X_{l\omega}^{\rm (int)}$ is derived from 
 the linearized Einstein equation
for axial modes as
%%%%%%%%%
\begin{equation}
\frac{d^2 X_{l\omega}^{\rm (int)}}{d r^{*2}} 
+ \left( \omega^2 - V_{l}^{\rm (int)} \right) X_{l\omega}^{\rm (int)} = 
0, 
\label{eq:inside_perturbation_equation}
\end{equation}
%%%%%%%%%
where the ``tortoise'' coordinate $r^{*}$  and the effective potential
$V_{l}^{\rm (int)}$ 
 are defined as 
%%%%%%%%%
\begin{eqnarray}
r^{*} &=& \int_{0}^{r} e^{- (\nu - \lambda)/2} dr,
\\
V_{l}^{\rm (int)} &=& e^{\nu} 
\left( \frac{l(l+1)}{r^2} - \frac{6M(r)}{r^3}
- 4 \pi (P - \rho) \right),
\label{eq:inside_perturbation_equation_potential}
\end{eqnarray}
%%%%%%%%%
respectively.

As for the boundary condition,
the wave function $X_{l\omega}^{\rm (int)}$ must be regular at the 
center of a star.
Expanding  $X_{l\omega}^{\rm (int)}$ 
around the center $r=0$, we find 
%%%%%%%%%
\begin{equation}
X_{l\omega}^{\rm (int)} = 
\eta_{l\omega} \left\{ r^{l+1} + \frac{1}{2(2l+3)} 
\left[ 4 \pi (l + 2)
\left( \frac{1}{3} (2l - 1) \rho_{\rm c} - P_{\rm c} \right)
- \omega^2 e^{- \nu_{\rm c}} \right] r^{l + 3} + O(r^{l+5}) \right\},
\label{eq:BC_X}
\end{equation}
%%%%%%%%%
where $\eta_{l\omega}$ is an arbitrary constant, and 
$\rho_{\rm c}$, $P_{\rm c}$, and $\nu_{\rm c}$ are
the central density, pressure, and metric functions, 
respectively.

Apart from a free parameter $\eta_{l\omega}$, we obtain a solution 
$X_{l\omega}^{\rm (int)}$  by integrating
Eq. (\ref{eq:inside_perturbation_equation}) 
with the boundary condition ($\ref{eq:BC_X}$).
Further details of the numerical techniques constructing the wave 
function
$X_{l\omega}^{\rm (int)}$ are given 
in the Appendix, subsection \ref{subsec:app_interior}.

%%%%%%%%%%%%%%%%%%%%%%%%
\subsection{Exterior region of the star}
\label{subsec:exterior}
%%%%%%%%%%%%%%%%%%%%%%%%

The background metric outside the star
is the  Schwarzschild spacetime, i.e. ,
 Eq. (\ref{eq:bgmetric}), with
%%%%%%%%%
\begin{eqnarray}
e^{\nu} = e^{-\lambda} = 1 - \frac{2 M}{r}.
\end{eqnarray}
%%%%%%%%%

As for the trajectory of a test particle, without loss of generality, we
can assume that the particle moves on the equatorial plane  ($\theta =
\pi / 2$).  Then, the equation of motion for the particle 
is described as 
%%%%%%%%%
\begin{eqnarray}
\frac{d \hat{T}}{d \tau} 
&=& 
\tilde{E} \left/ \left(1 - \frac{2M}{\hat{R}} \right) \right., 
\\
\left( \frac{d \hat{R}}{d \tau} \right)^2 
&=& 
\tilde{E}^2 - 
\left( 1 - \frac{2M}{\hat{R}} \right) 
\left( 1 + \frac{\tilde{L}^2}{\hat{R}^2} \right), 
\label{eq:test_particle_geodesics_r} 
\\
\hat{\Theta} 
&=&
\frac{\pi}{2},
\\
\frac{d \hat{\Phi}}{d \tau} &=& \frac{\tilde{L}}{\hat{R}^2},
\end{eqnarray}
%%%%%%%%%
where the particle trajectory is 
$Z^\mu= {\bf (} \hat{T}(\tau),\hat{R}(\tau),\hat{\Theta}(\tau),\hat{\Phi}(\tau) {\bf)}$,
and 
 $\mu$, $\tilde{E}$, and $\tilde{L}$ are the mass, 
the  normalized energy, and
the  normalized orbital angular momentum of the particle, respectively, 
with 
$\tilde{E} = E / \mu$ and $\tilde{L} = L / \mu$. 
The normalized effective  potential energy $\tilde{V}^{\rm
(particle)}(r)$ of the particle is given as
%%%%%%%%%
\begin{equation}
\tilde{V}^{\rm (particle)}(r) = 
\sqrt{\left( 1 - \frac{2M}{r} \right) 
\left( 1 + \frac{\tilde{L}^2}{r^2} \right)}.
\label{eq:potential_energy} 
\end{equation}
%%%%%%%%%
In this paper, since we only consider the motion of a test particle 
scattered by a spherical star,
 the particle  orbit is  unbounded, i.e.,
%%%%%%%%%
\begin{equation}
|\tilde{L}| / M > 4 , 
\qquad 1 \le \tilde{E} < 
\tilde{V}_{\rm max}^{\rm (particle)},
\nonumber
\end{equation}
%%%%%%%%%
where $\tilde{V}_{\rm max}^{\rm (particle)}$ is the maximum value of the
potential.

The perturbed metric outside a star only comes from
an oscillation of spacetime, that is, gravitational waves.
Then, similar to the interior region, the linearized Einstein equations
for the metric $h_{1,lm}$ and $ 
h_{0,lm}$  defined as Eq. (\ref{metric})
are reduced to a single wave equation
for a new wave function $X_{lm\omega}^{\rm (ext)}$ defined 
by 
%%%%%%%%%
\begin{eqnarray}
h_{1,lm \omega} &=& e^{\lambda} r X_{lm\omega}^{\rm (ext)}, \\
h_{0,lm \omega} &=& \frac{i}{\omega} e^{-\lambda} \frac{d}{dr} 
\left( r X_{lm\omega}^{\rm (ext)} \right)
- \frac{8 \pi i}{\omega} \frac{r^2 e^{-\lambda}}{\sqrt{2n(n+1)}} 
D_{lm}(\omega, r), 
\end{eqnarray}
%%%%%%%%%
where  $n \equiv (l - 1)(l + 2) / 2$ and   $D_{lm}(\omega, r)$ 
is one of the source terms given below.

From the linearized vacuum Einstein equations, we find the wave equation 
for
$X_{lm\omega}^{\rm (ext)}$ (Regge-Wheeler equation)\cite{RW} as 
%%%%%%%%%
\begin{eqnarray}
\frac{d^2 X_{lm\omega}^{\rm (ext)}}{d r^{*2}} 
+ \left( \omega^2 - V_{l}^{\rm (ext)} \right) X_{lm\omega}^{\rm (ext)} 
= S_{lm\omega}^{\rm (ext)},
\label{eq:outside_perturbation_equation}
\end{eqnarray}
%%%%%%%%%
where  $r^{*}$ is the 
tortoise coordinate 
defined by 
%%%%%%%%%
\begin{equation}
r^{*} = r + 2M \ln \left( \frac{r}{2M} - 1 \right), 
\end{equation}
$V_{l}^{\rm (ext)}$ is 
Regge-Wheeler potential given by
%%%%%%%%%
\begin{eqnarray}
V_{l}^{\rm (ext)} & = &
e^{- \lambda} \left( \frac{l(l+1)}{r^2} - \frac{6M}{r^3} \right), 
\label{eq:outside_perturbation_equation_potential}
\label{RW_potential}
\end{eqnarray}
%%%%%%%%%
and $S_{lm\omega}^{\rm (ext)}$  is  the source term,
which is described by the  energy-momentum tensor of
a test particle expanded by tensor harmonics\cite{Zerilli} 
as 
\begin{eqnarray}
S_{lm\omega}^{\rm (ext)} &=& 
8 \pi \left[ \frac{e^{-2 \lambda}}{\sqrt{n+1}} Q_{lm}(\omega, r)
+ \frac{r e^{- \lambda}}{\sqrt{2n(n+1)}} \frac{d}{dr}  \left[
e^{-\lambda}  D_{lm}(\omega, r) \right] \right],
\label{source}
\end{eqnarray}
with $Q_{lm}(t,r)$ and
$D_{lm}(t,r)$ being
%%%%%%%%%
\begin{eqnarray}
Q_{lm}(t,r) &=& 
\frac{1}{\sqrt{n + 1}} \frac{\mu}{r} e^{\lambda} \int_{-\infty}^{\infty}
d \tau \frac{d \hat{R}}{d \tau} \delta {\bf (} t - \hat{T}(\tau) {\bf )} \delta {\bf (} r  -
\hat{R}(\tau) {\bf )} \nonumber 
\\
& & \quad \times \left[ \frac{1}{\sin \hat{\Theta}} \frac{\partial 
\bar{Y}_{lm}}{\partial \hat{\Phi}} \frac{d \hat{\Theta}}{d \tau} - \sin
\hat{\Theta} \frac{\partial \bar{Y}_{lm}}{\partial \hat{\Theta}} \frac{d
\hat{\Phi}}{d \tau} \right], 
\label{eq:def_Q} \\
D_{lm}(t,r) &=& 
- \frac{1}{\sqrt{2n(n+1)}} \mu \int_{-\infty}^{\infty}
d \tau \delta {\bf (} t - \hat{T}(\tau) {\bf )} \delta {\bf (} r - \hat{R}(\tau) {\bf )} 
\nonumber \\
& & 
\quad \times \Biggl\{ \frac{1}{2} 
\left[ \left( \frac{d \hat{\Theta}}{d \tau} \right)^2 
- \sin^2 \hat{\Theta} 
\left( \frac{d \hat{\Phi}}{d \tau}\right)^2 \right] 
\frac{1}{\sin \hat{\Theta}} \bar{X}_{lm} (\hat{\Omega}) 
\nonumber 
\\
& & 
\qquad \qquad \qquad \qquad \qquad - 
\sin \hat{\Theta} 
\frac{d \hat{\Theta}}{d \tau} 
\frac{d \hat{\Phi}}{d \tau} 
\bar{W}_{lm}(\hat{\Omega})
\Biggr\}.
\label{eq:def_D}
\end{eqnarray}
%%%%%%%%%
$X_{lm}$, $W_{lm}$ are 
the tensorial part of a tensor harmonics defined by
%%%%%%%%%
\begin{eqnarray}
X_{lm} &=& 2 \frac{\partial}{\partial \phi} \left( 
\frac{\partial}{\partial 
\theta} - \cot \theta \right) Y_{lm}, \\
W_{lm} &=& \left( \frac{\partial^2}{\partial \theta^2} 
- \cot \theta \frac{\partial}{\partial \theta} 
- \frac{1}{\sin^2 \theta} \frac{\partial^2}{\partial \phi^2}  \right)
Y_{lm},
\end{eqnarray}
%%%%%%%%%
and the overbar denotes the complex conjugate. 
%%%%%%%%%

The boundary condition outside the star should be imposed such that
there is no incoming wave at infinity, 
%%%%%%%%%
\begin{equation}
X_{lm\omega}^{\rm (ext)} \to A_{lm\omega} e^{i \omega r^{*}} 
\qquad (r^{*} \to \infty),
\label{eq:BC_infty}
\end{equation}
%%%%%%%%%
where $A_{lm\omega}$ is the amplitude of an outgoing wave at infinity.
The details of constructing the wave function 
$X_{lm\omega}^{\rm (ext)}$
are given in the Appendix, subsections \ref{subsec:app_exterior} and 
\ref{subsec:app_turning_point}.

%%%%%%%%%%%%%%%%%%%%%%%%
\subsection{Matching of two wave functions and gravitational waves}
%%%%%%%%%%%%%%%%%%%%%%%%

We have to combine two wave functions,
the interior wave function $X_{l\omega}^{\rm (int)}$ and 
 the exterior one $X_{lm\omega}^{\rm (ext)}$,
at the surface of the star.
The matching condition is that 
the wave function  must be continuous and smooth 
at the surface.
For the case of the axial mode, since the wave function inside the star
does not couple to perturbations of the matter fluid,  the condition
turns out to be very simple  as
%%%%%%%%%
\begin{eqnarray}
X_{l\omega}^{\rm (int)}(R^{*}) &=& X_{lm\omega}^{\rm (ext)}(R^{*}), 
\label{eq:S_condition1} \\
 \frac{dX_{l\omega}^{\rm (int)}}{dr^{*}} \left( R^{*}\right) &=& 
 \frac{dX_{lm\omega}^{\rm (ext)}}{dr^{*}}\left( R^{*}\right),
\label{eq:S_condition2}
\end{eqnarray}
%%%%%%%%%
where $R^{*}$ is the radius of the star by use of the ``tortoise" coordinate.

Then  we calculate the amplitude of the gravitational waves $A_{lm\omega}$ as
follows:  We first integrate Eqs. (\ref{eq:inside_perturbation_equation}) and
(\ref{eq:outside_perturbation_equation}) 
using their boundary conditions, 
Eqs. (\ref{eq:BC_X}) and (\ref{eq:BC_infty}).
Next, we connect those solutions at the surface of the star using 
Eqs. (\ref{eq:S_condition1}) and (\ref{eq:S_condition2}), which 
determine  two unknown coefficients $\eta_{l\omega}$ and $A_{lm\omega}$.

From the wave function obtained as above, we obtain some information about 
gravitational waves emitted by a scattered particle.
 When we discuss the
gravitational waves, we usually  decompose them into two modes:
$+$ and $\times$ modes.
Then, 
we have to transform our coordinate system  from the Regge-Wheeler gauge
to the transverse-traceless one, which  metric near infinity is 
described as
%%%%%%%%%%
\begin{equation}
ds^{2} = -dt^{2} + dr^{2} + r^{2} (1 + h_{+}) d\theta^{2} + 
r^{2} \sin^{2}\theta (1 - h_{+}) d\phi^{2} + 
r^{2} \sin\theta  h_{\times}  d\theta d\phi.
\end{equation}
%%%%%%%%%
Then the axial mode
in transverse-traceless gauge at infinity turns out to be  
%%%%%%%%%%
\begin{eqnarray}
h_{+} \pm i h_{\times} =
\mp \frac{1}{\pi r} \int_{- \infty}^{\infty} \frac{d
\omega}{\omega } 
\sum_{l=2}^{\infty} 
\sum_{m=-l}^{l} A_{lm \omega}
 \sqrt{n(n+1)} e^{i \omega (r^{*} - t)} {}_{\pm 2} Y_{lm}
(\theta, \phi), 
\end{eqnarray}
%%%%%%%%%
where the spin-weighted spherical harmonics ${}_{\pm 2} Y_{lm}$ is defined by
%%%%%%%%%%
\begin{eqnarray}
{}_{\pm 2} Y_{lm} (\theta, \phi) = \frac{1}{2 \sqrt{n(n+1)}} 
\left( 
W_{lm} \pm \frac{i}{\sin \theta} X_{lm}
\right).
\end{eqnarray}
%%%%%%%%%

The energy-momentum tensor 
of the gravitational waves at infinity is given by 
%%%%%%%%%
\begin{eqnarray}
T_{\mu \nu}^{\rm (GW)} = \frac{1}{32 \pi} 
\left< 
{\rm Re} (\nabla_{\mu} h_{\alpha \beta}^{\rm (TT)}) 
{\rm Re} (\nabla_{\nu} h^{\alpha \beta {\rm (TT)}}) \right>,
\label{eq:EMT_GW}
\end{eqnarray}
%%%%%%%%%
in the transverse-traceless gauge, where the angular brackets denote an appropriate
average.

From this definition, we find  the energy spectrum $d E_{\rm GW} / d \omega$ 
 of gravitational waves at infinity for the axial mode
 as 
%%%%%%%%%
\begin{equation}
\frac{dE_{\rm GW}}{d \omega} =
\frac{1}{32 \pi} \sum_{l=2}^{\infty} 
\sum_{m=-l}^{l} l (l + 1) (l - 1) (l + 2) 
\left|A_{lm \omega} \right|^2.
\end{equation}
%%%%%%%%%

%%%%%%%%%%%%%%%%%%%%%%%%%%%%%%%%%%%%%%%%%%%%%%%%%
\section{Gravitational waves from a scattered test particle}
\label{sec:GW}
%%%%%%%%%%%%%%%%%%%%%%%%%%%%%%%%%%%%%%%%%%%%%%%%%

%%%%%%%%%%%%%%%%%%%%%%%%
\subsection{Uniform density star}
%%%%%%%%%%%%%%%%%%%%%%%%

In this subsection, we consider  a uniform density star.
We restrict our analysis only to the case of
the  $l=2$ mode, just for simplicity.
We find that features of the energy spectrum of emitted gravitational
waves can be classified into two cases, 
which mainly depend on the shape of the effective potential.

\begin{figure}
\begin{center}
\leavevmode
\epsfysize=220pt
\epsfbox{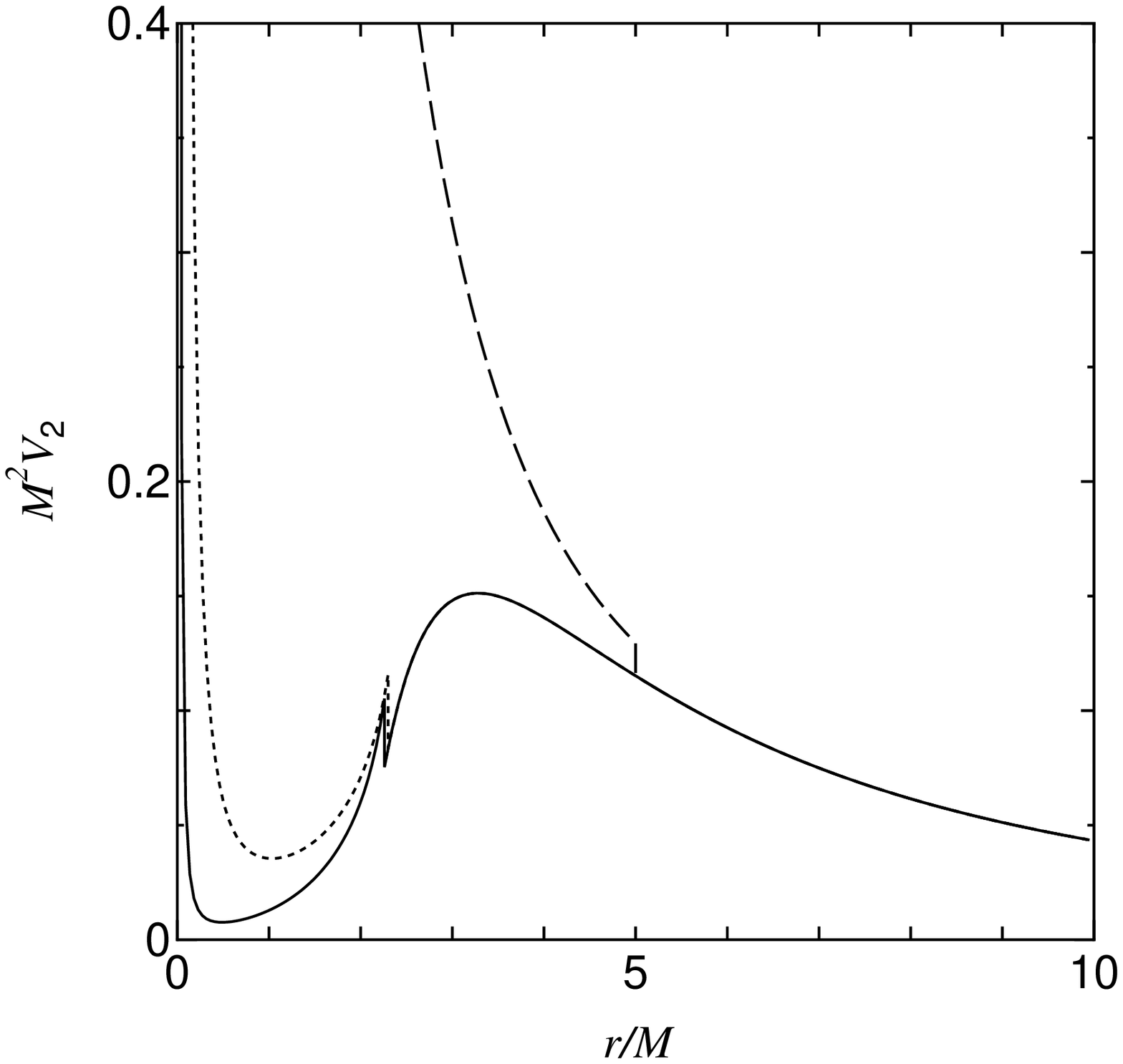}
\end{center}
\caption{
Effective potentials of a uniform density
star for the $l=2$ mode. Solid, dotted, and dashed lines
correspond to stellar models with radius 
$R=2.26M$, $2.3M$, and $5.0M$, respectively.
For an ultracompact star 
($R < 3M$), 
the potential has a minimum, which traps gravitational 
waves emitted by a test particle. 
}
\label{fig:potential}
\end{figure}

Figure \ref{fig:potential} shows the effective potential 
[Eqs. (\ref{eq:inside_perturbation_equation_potential}) and 
(\ref{eq:outside_perturbation_equation_potential}) ]
 for a uniform density star 
with three different compactness parameters $R/M$.
The effective potential has the following three features.
First, the Regge-Wheeler potential 
[Eq. (\ref{eq:outside_perturbation_equation_potential})]
has a maximum point near $r \sim 3 M$
(accurately at $r = 3.28 M$ for $l = 2$).
Second, the potential  is discontinuous at the  surface because
of a discontinuity of the energy density of the star  at the
surface. Finally,  a potential minimum appears inside the star,
if the star is   ultracompact  (for the case of $R \lesssim 3M$).

According to the last feature,
we study two cases separately.
The first stellar type is the case that the surface
is outside the potential barrier ($R \gtrsim 3  M $).
For a usual  neutron star,  $R \sim 5M$, and then 
the first type is most likely a realistic astronomical object.
The second stellar type is the case that 
the surface is inside the potential barrier ($R  \lesssim 3  M$),
which may be an exotic star, if it exists.

\begin{figure}
\begin{center}
\leavevmode
\epsfysize=220pt
\epsfbox{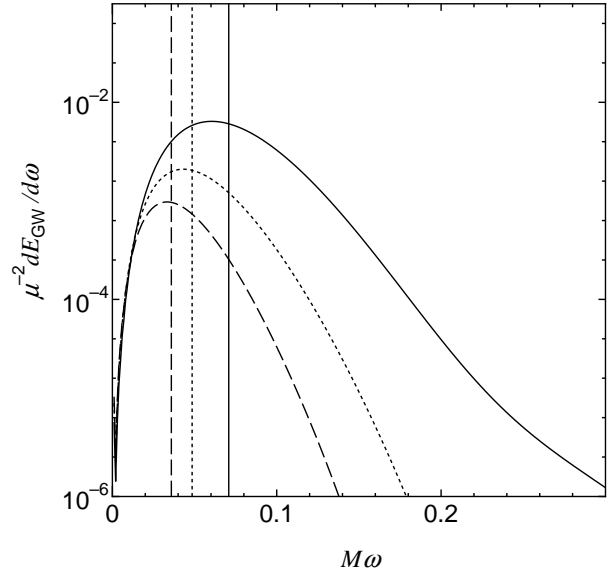}
\end{center}
\caption{
The energy spectrum ($l=2$ mode) of
gravitational waves from a test particle with  $E  =
1.01\mu$ scattered by a uniform density star with 
$R=5.0M$.
Solid, dotted, and dashed lines show the cases of
the angular momentum 
$L=4.5\mu M$, $5.0\mu M$, and  $5.5\mu M$,
respectively.
Longitudinal lines represent the frequency of a test particle at the
turning point.
Peaks correspond to orbital frequencies at the turning point of
a test particle.
}
\label{fig:rm5}
\end{figure}

\begin{figure}
\begin{center}
\leavevmode
\epsfysize=220pt
\epsfbox{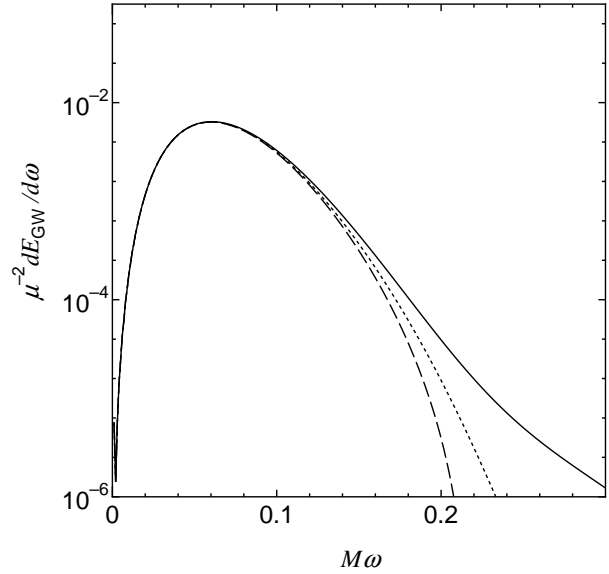}
\end{center}
\caption{The energy spectrum  ($l=2$ mode) of
gravitational waves for  a uniform density star with
$R=5.0M$  and for a Schwarzschild black hole.
We choose the energy and angular momentum of the test particle 
as $(E, L)=(1.01\mu,4.5\mu M)$.
Solid, dotted, and dashed lines correspond to  the cases of the uniform
density star with 
$R=5.0 M$, $4.0 M$, and of
a Schwarzschild black hole, respectively.
The peak in the spectrum coincides each other  because of the same
trajectory of the test particle.
}
\label{fig:rm5bh}
\end{figure}

First we show our result for the first type.
For the case of $R  = 5.0 M $, we show the energy spectra of emitted
gravitational waves in Fig. \ref{fig:rm5}.
In this figure, there exists only a single peak in each spectrum.
This peak corresponds to 
the orbital frequency  of a test particle at the turning point, i.e.,
periastron. This means that most gravitational waves are
emitted around the turning point, where 
the effect of the gravitational field is the strongest.
No other feature is found  in this spectrum.
In fact, fixing  the energy $E$ and 
orbital angular momentum $L$, 
the spectrum depends less on the compactness parameter  $R/M$ (see
Fig. \ref{fig:rm5bh}). For a black hole background, we find quite similar
behaviors in the energy spectrum (Fig. \ref{fig:rm5bh}), except for the high frequency
region, where  the difference of the boundary conditions is crucial.
We also find that  the spectrum approaches
closer to that of a black hole as the star gets smaller.

\begin{figure}
\begin{center}
\leavevmode
\epsfysize=220pt
\epsfbox{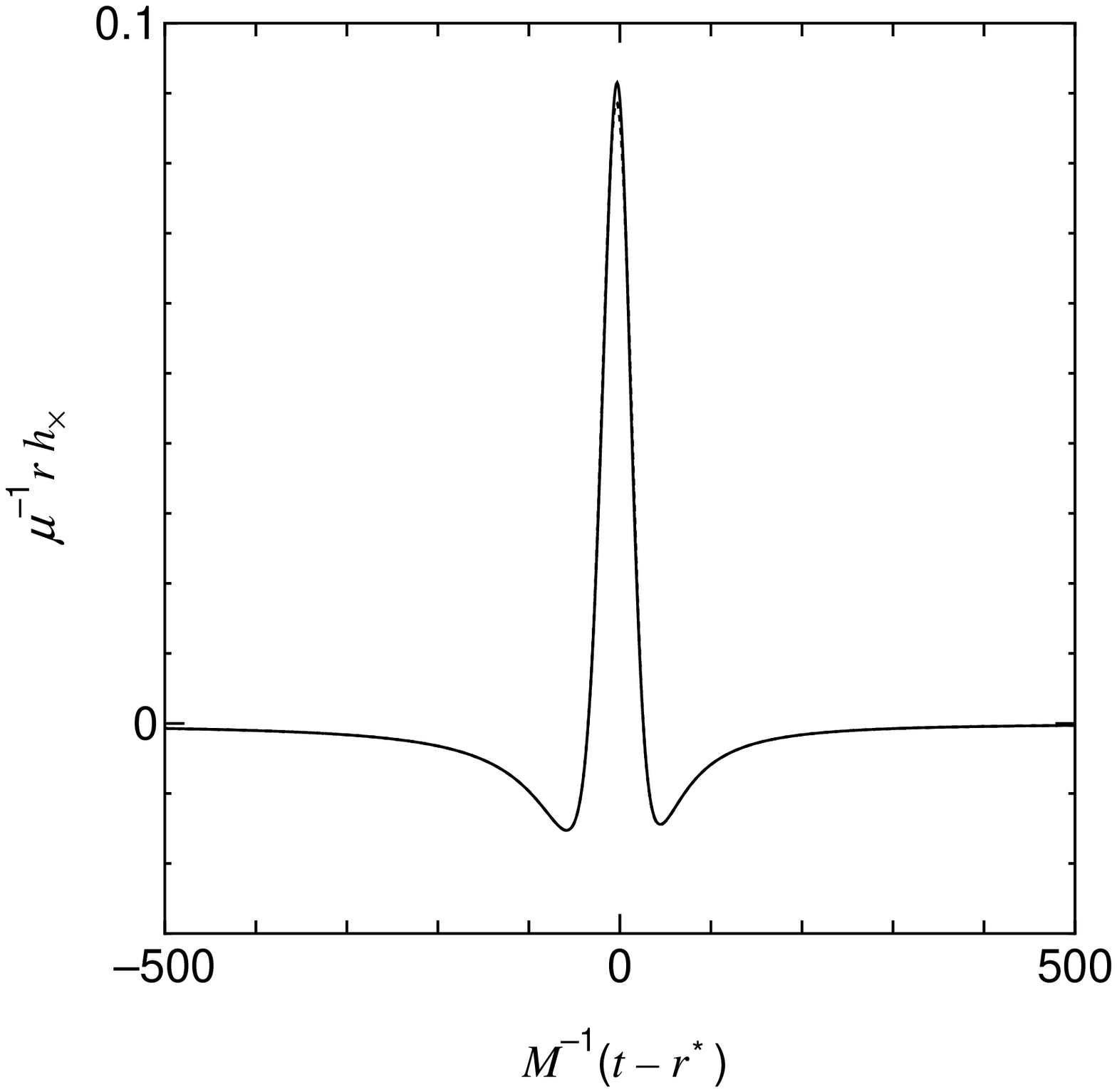}
\end{center}
\caption{
Waveforms observed in the direction of
$\theta =
\pi/2$, $\phi = 0$ ($l=2$
mode) for a uniform density star with 
$R=5.0 M$ (solid line) and for a Schwarzschild black hole (dotted line).
We set the energy and angular momentum of the test particle 
as $(E,L)=(1.01 \mu,4.5 \mu M)$.
In this case, the $+$ mode cancels out in the waveform.
}
\label{fig:wf5}
\end{figure}

We also calculate the waveform for the case of $R = 5.0 M$ and black holes, 
respectively.
We see that a burst wave due to the encounter of a test particle emerges for each case.
As the same as the energy spectrum, the difference between these waveform is very little
(see Fig. \ref{fig:wf5})

 Therefore, for the first stellar type of $R >3M$, 
the energy spectrum
largely depends on the trajectory, 
but not on the background object.

\begin{figure}
\begin{center}
\leavevmode
\epsfysize=220pt
\epsfbox{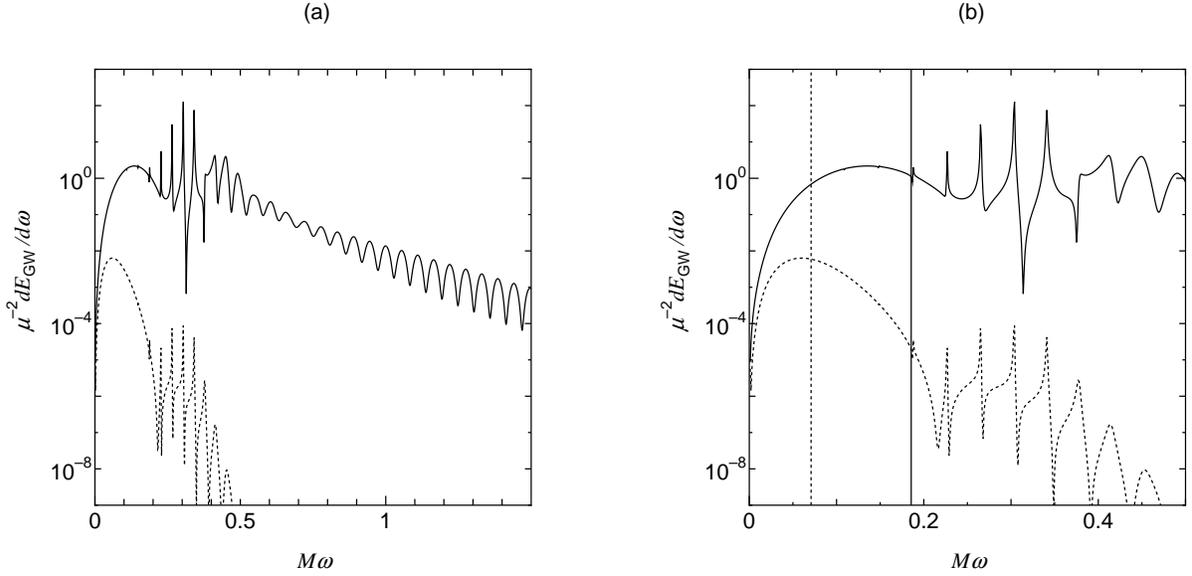}
\end{center}
\caption{
The energy spectrum ($l=2$ mode) of 
gravitational waves for the case of a uniform density
star with
$R=2.26 M$.
Solid and dotted lines correspond to 
$(E, L)=(2.38 \mu,12.0 \mu M)$ and $(1.01 \mu,4.5 \mu M)$,
respectively.
(b) is an enlargement of (a) to show the resonant peaks in 
detail. 
Longitudinal lines represent the frequency of a test particle at the
turning point.
We find  some periodic  peaks,
which never appear in Fig. 2. Those small periodic peaks are classified into two
types: one for $\omega < \omega_{\rm max}$ and the other for $\omega >
\omega_{\rm max}$, where $\omega_{\rm max}$ corresponds to the maximum
energy of the effective potential in Fig. 1. 
}
\label{fig:rm2.26}
\end{figure}

Next we discuss an ultracompact star, which shows much more
various structure in the spectrum. 
In this case, a minimum point of the
effective potential appears inside the star. As for the
trajectory of a test particle, we have two possibilities:  one is
that the particle gets into the inside region beyond 
 the potential barrier  and  the other is that one passes  through 
outside the barrier.
Since the particle approaches closer to the star, it feels a stronger
gravitational field. As a result, the excitation by the particle is more
efficient and the emitted gravitational waves are enhanced.
As a
numerical example, we show those energy spectra for the model with
radius $R =2.26M$ with 
 the turning points
$\hat{R}_{\rm min} = 3.145M$  and  $ 6.628M $ in Fig. \ref{fig:rm2.26}.  

Apart from a
global  maximum, which appears in the first type model as well and
corresponds to  the orbital frequency  of a test particle at the turning
point,  we also find many small  periodic peaks in Fig. \ref{fig:rm2.26}.  This
microscopic peak structure is classified into two types, which  are
divided at the frequency of the maximum 
$\omega_{\rm max} \equiv \sqrt{V_{l \ {\rm max}}^{\rm (ext)}}$ 
($\simeq 0.389 M^{-1}$ for $l = 2$) of the
effective potential. In the lower frequency region ($\omega <
\omega_{\rm max}$),  small and sharp peaks correspond to the
quasinormal modes, as we will  show it next. For the higher frequency
region ($\omega > \omega_{\rm max}$),  we find small periodic peaks,
which may appear because of a  resonance between two potential barriers. 
We will also discuss this later.

In order to analyze our numerical results, we first have to know
the axial quasinormal modes.
Our method to calculate a quasinormal mode is based on 
the continued fraction expansion method which was originally 
used for black holes by Leaver \cite{Leaver}, and adopted for 
the polar mode of a spherical star by
Leins, Nollert, and Soffel \cite{LNS}.
We apply this method to axial modes as well.
We show our results  in Table \ref{tab:QNM_Urm2.26}  with those of
Kokkotas\cite{Kokkotas} whose method is different from ours.  The
table shows that both results agree quite well.  

\begin{figure}
\begin{center}
\leavevmode
\epsfysize=220pt
\epsfbox{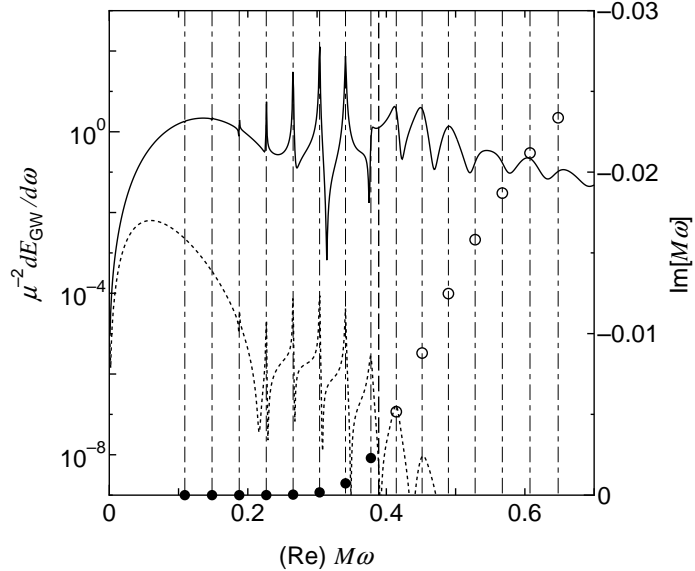}
\end{center}
\caption{
The energy spectrum  ($l=2$ mode) of 
gravitational waves  and axial ``quasinormal" modes for
the case of a  uniform density star with
$R=2.26 M$.
Solid  and dotted lines correspond to  
$(E, L )=(2.38 \mu,12.0\mu M)$ and $(1.01 \mu,4.5\mu M)$,
respectively, which are the  same those in  Fig. 5.
A longitude dash line represents the frequency of the maximum
$\omega_{\rm max}$ of the effective potential.
Solid ($\omega < \omega_{\rm max}$) and open ($\omega > \omega_{\rm
max}$) circles denote axial ``quasinormal" modes of the  uniform density
star with $R=2.26 M$.
We find that resonant periodic peaks for $\omega < \omega_{\rm max}$
agree quite well with those quasinormal modes, which are excited by a
scattered test particle. 
However, small periodic peaks for $\omega > \omega_{\rm max}$
seem not coincide  exactly with ``quasinormal" modes.
}
\label{fig:rm2.26qnm}
\end{figure}

In Fig. \ref{fig:rm2.26qnm},  we present both the quasinormal mode and energy spectrum.
As seen from Fig. \ref{fig:rm2.26qnm}, axial ``quasinormal" modes are classified into 
two:  the lower frequency mode ($\omega < \omega_{\rm max}$), which
imaginary part is very small ${\rm Im} (\omega) < O(10^{-8})$, and the
higher frequency mode ($\omega >
\omega_{\rm max}$), which imaginary part is rather large
$(10^{-8}) < {\rm Im} (\omega)< O(1)$.
The reason is very simple. For a  frequency lower than 
$\omega_{\rm max}$, we can have a quasibound state.  Since 
trapped waves tunnel through the potential barrier, the energy will
decrease. It is similar to the quasinormal mode of a black hole. 
Then, we may call it the axial quasinormal mode as well.  However,
if the frequency is higher than $\omega_{\rm max}$, we have no bound
state.  Nevertheless, we can find the ``quasinormal" mode, by which we
mean a wave solution with the same boundary condition as that of
the conventional quasinormal mode, which is the outgoing wave
condition. In the present case, although we find such wave
solutions, those waves decay very quickly because the frequency
is higher than the potential barrier.  This is the reason why we
find a large imaginary part of the mode frequency. The solutions which
satisfy the boundary condition appear periodically as seen from Fig. \ref{fig:rm2.26qnm}.

\begin{figure}
\begin{center}
\leavevmode
\epsfysize=220pt
\epsfbox{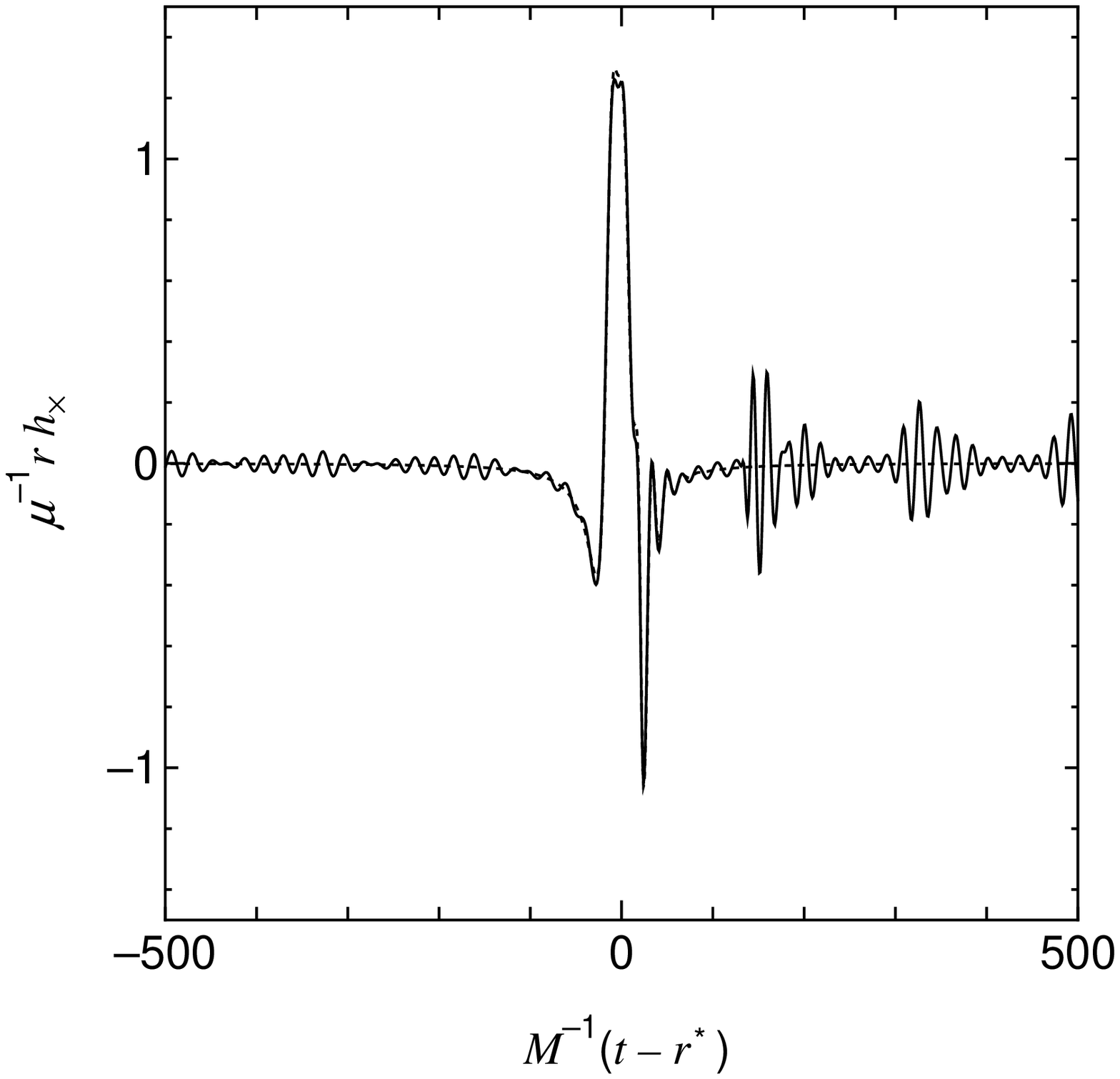}
\end{center}
\caption{
The waveform observed in the direction of
$\theta =
\pi/2$, $\phi = 0$ ($l=2$
mode) for a uniform density star with 
$R=2.26 M$ (solid line) and for a Schwarzschild black hole (dotted line).
We set the energy and angular momentum of the test particle 
as $(E,L)=(2.38 \mu,12.0 \mu M)$.
In this case, the $+$ mode cancels out in the waveform.
}
\label{fig:wf2.26}
\end{figure}

We find that the frequencies of these quasinormal modes 
coincide with those of spectrum peaks in the frequency region
lower  than  $\omega_{\rm max}$.
This is because a test particle excites those quasinormal modes.
In fact, when we see the waveform (Fig. \ref{fig:wf2.26}), the wave consists of  
two parts: large one, which corresponds to the burst wave by the
encounter by a test particle, and the other part with several
oscillations, which comes from exited quasinormal modes in the
spectrum.
Actually the waveform in the region after $t - r^{*} \sim 100 M$ in Fig. \ref{fig:wf2.26}
is similar to Kokkotas's result for a Gaussian pulse wave \cite{Kokkotas_proceeding}.

On the other hand, in a frequency region higher than  
$\omega_{\rm max}$,  although  many periodic peaks appear, those
positions are not exactly the same as the frequencies of
``quasinormal" modes. We may understand this fact as follows.
Even if the wave frequency $\omega$ is higher than the potential 
barrier, the reflection coefficient does not vanish and  oscillates
with respect to
$\omega$, which period is determined by the potential form.
In the present case, we have also an infinite wall near the center 
of a star.  We then expect  that small periodic peaks appearing in
the spectrum are due to an interference between the incoming and
outgoing waves reflected by two potential barriers. Since these
waves decay very quickly as mentioned above, the process is quite
dynamical, and then  the peaks in the spectrum do not coincide with
the frequencies of ``quasinormal" modes.

\begin{figure}
\begin{center}
\leavevmode
\epsfysize=220pt
\epsfbox{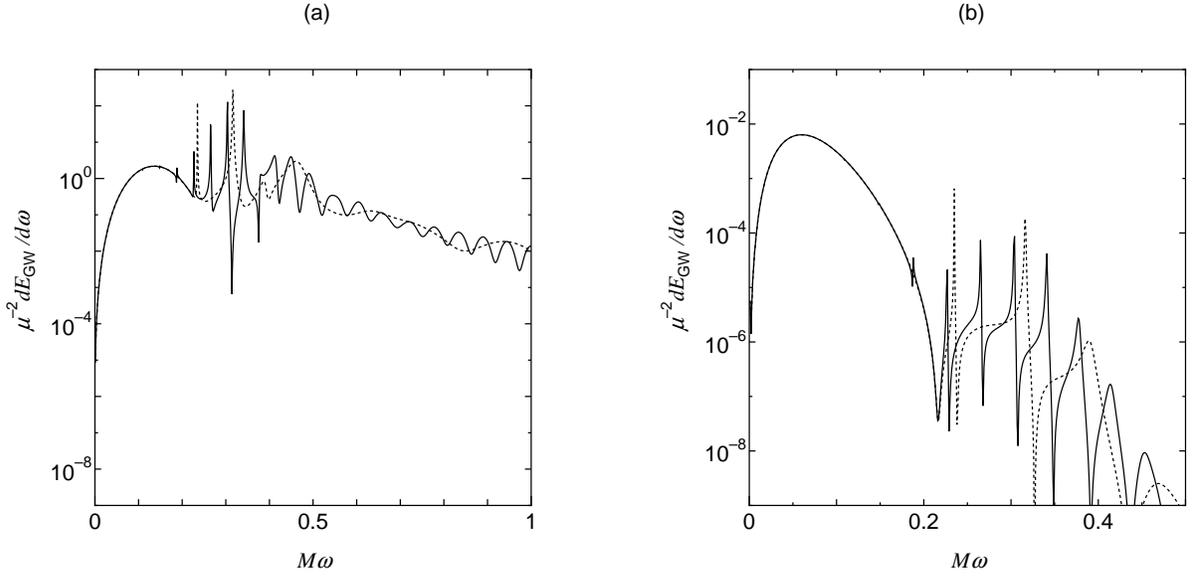}
\end{center}
\caption{
The energy spectrum  ($l=2$ mode)  of
gravitational waves for uniform density stars with  two
different compactness parameters, 
$R=2.26 M$ (solid line) and $2.3 M$ (dotted line).
We set the energy and angular momentum of the test particle 
as (a) $(E,L)=(2.38 \mu,12.0 \mu M)$ and 
(b) $(E,L)=(1.01 \mu,4.5 \mu M)$.
Although global peaks coincide with each other because of the same
trajectory of the particle, resonant periodic peaks are different
because of the different compactness.  This may show us how to
distinguish the compactness of a star by use of observational data.
}
\label{fig:rm2.26-2.3}
\end{figure}

In Fig. \ref{fig:rm2.26-2.3}, we show the energy spectra for 
two stellar models with $R 
  = 2.26M$, and  $2.3M$, which 
correspond to the cases that the turning point is inside and 
outside the barrier, respectively. For fixed energy $E $ and 
angular momentum $L$, 
the global peaks corresponding to the orbital frequency at the 
turning point  agree with each other.
On the other hand, the other small periodic  peaks from the  excited
quasinormal modes  do not agree with and rather depend on the 
compactness parameter $R/M$. Consequently, from the analysis of the
energy spectrum of axial modes,  we may determine the compactness
parameter $R/M$ of an ultracompact star.

\begin{figure}
\begin{center}
\leavevmode
\epsfysize=220pt
\epsfbox{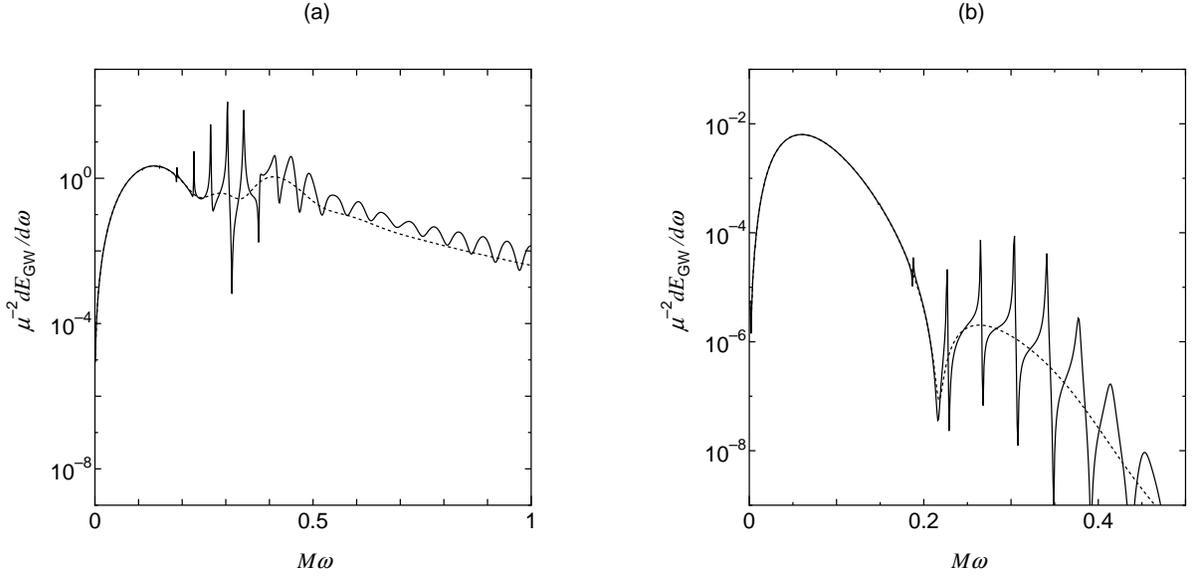}
\end{center}
\caption{
The energy spectra   ($l=2$ mode)  of
gravitational waves for  a uniform density star with
$R=2.26 M$ (solid line) and for  a Schwarzschild black
hole (dotted line). We set the energy and angular
momentum of the test particle  as (a) $(E,L)=(2.38
\mu,12.0 \mu M)$ and  (b) $(E,L)=(1.01 \mu,4.5 \mu M)$.
This difference may enable us to distinguish an ultracompact star from a
black hole.
}
\label{fig:rm2.26bh}
\end{figure}

We can also compare a black hole and an ultracompact star.
In Fig. \ref{fig:rm2.26bh}, we show  
 the energy spectra both  for a black hole 
\cite{ON} and an ultracompact star with $R 
 = 2.26 M$.
From  Fig. \ref{fig:rm2.26bh}, we see that  even for the case of a Schwarzschild 
black hole, the global peak due to the orbital motion of the
particle coincides with that for the star.
However, unlike the stellar model, 
no small periodic peaks appear in the case of a black hole.
In fact for a black hole, there is no peak after $t - r^{*} \sim 100 M$ (see Fig. 7).
This is because the
effective potential of the star  has an infinite potential wall 
near the center, which guarantees the existence of quasinormal 
modes, while that of a Schwarzschild black hole  has no
such wall,  and incoming waves fall entirely into the event 
horizon 
$(r_H  = 2 M )$.
Hence, we can also distinguish a background object, i.e., 
whether it is an ultracompact star or a black hole, analyzing only
axial modes.

As for the total energy of emitted gravitational waves $E_{\rm GW}$, 
we find
%%%%%%%%%%%%%%
\begin{eqnarray}
E_{\rm GW} = \epsilon \ \frac{\mu^2}{M}, 
\end{eqnarray}
%%%%%%%%%%%%%%
with $\epsilon = 2.501 \times 10^{-2}$ and $3.541 \times 10^{-2}$ 
for $R = 5M$ and $4M$, respectively, 
and for $E = 2 \mu$ and $L = 12 \mu M$.
The efficiency $\epsilon$ increases monotonically as 
the compactness gets smaller, and seems to end up with that 
for Schwarzschild black holes ($\epsilon = 5.687 \times 10^{-2}$).
For $E = 1.01 \mu$ and $L = 4.5 \mu M$, the efficiency, which is
$\epsilon \simeq 2.3 \times 10^{-4}$, 
little depends
on $R / M$ up to the case of a black hole.

%%%%%%%%%%%%%%%%%%%%%%%%
\subsection{Polytropic star}
%%%%%%%%%%%%%%%%%%%%%%%%

From the analysis of the energy spectrum for a uniform  density star,
we find that the spectrum strongly depends on the  shape  of
effective potential. Therefore, it may be important to see  the
dependence of  the equation of state on the energy spectrum.
Then, in this section, we study a polytropic star.
Just for simplicity, we use the Newtonian polytropic equation of 
state as
%%%%%%%%%
\begin{eqnarray}
P = K \rho^{1 + 1 / n},
\end{eqnarray}
%%%%%%%%%
where $n$ is a polytropic index and $K$ is a constant value.

\begin{figure}
\begin{center}
\leavevmode
\epsfysize=220pt
\epsfbox{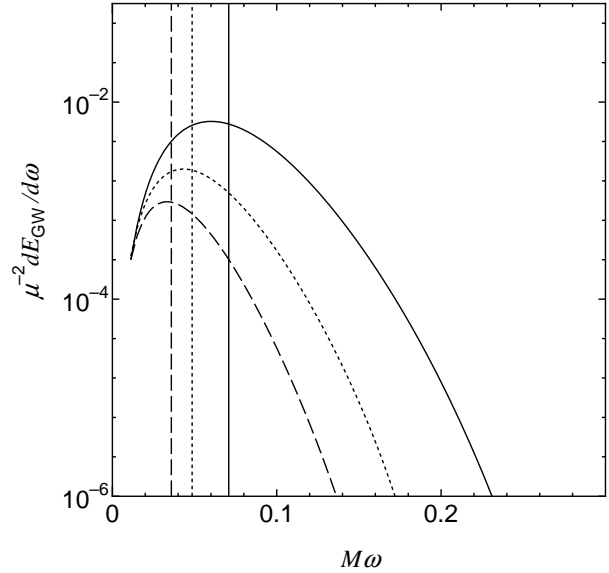}
\end{center}
\caption{
The energy spectrum   ($l=2$ mode) of
gravitational waves for the case of the polytropic star
with a large compactness parameter. We set the
polytropic index $n = 1$, the center of the density 
$\rho_{\rm c} = 3.0 \times 10^{15} $ g / cm$^{3}$, and the energy
of the test particle as $E = 1.01 \mu$. The solid, dotted, and
dashed lines correspond to the angular momentum of the test particle 
$L=4.5 \mu M$, $5.0 \mu M$, and $5.5 \mu M$,
respectively.
Longitudinal lines represent the frequency of a test particle at the
turning point.
This figure shows the same feature as Fig. 2.
}
\label{fig:polytropic1.0}
\end{figure}

First,  we consider the first type stellar model with 
$R \gtrsim 3M$.
We set 
the polytropic index $n = 1$
and the constant $K = 100 \ {\rm km^2}$, and  choose the  central
energy density  
$\rho_{\rm c} = 3.0 \times 10^{15} \ {\rm g / cm^{3}}$.
This choice gives a stellar model with mass $M = 1.267 
M_{\odot}$  and radius $R = 8.862 \ {\rm km}$.
The compact parameter is  $R/M 
 =  4.739$, and then 
the surface locates outside the peak of the Regge-Wheeler  potential.
We show the energy spectra of gravitational waves from  a
scattering test particle in Fig. \ref{fig:polytropic1.0}. We find the same feature  as
that for the uniform density star with 
$R \gtrsim 3 M$, i.e.,  there is only a single global peak which 
frequency corresponds to 
the orbital one of the particle at the turning point.

\begin{figure}
\begin{center}
\leavevmode
\epsfysize=250pt
\epsfbox{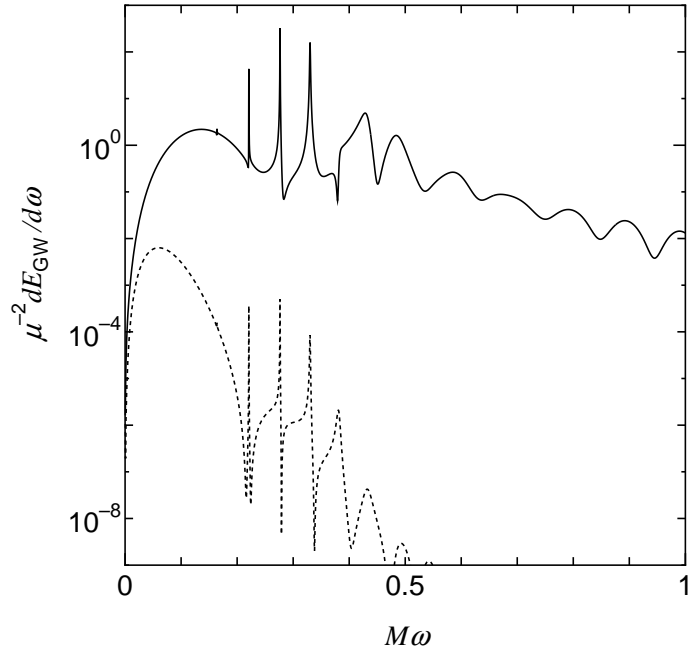}
\end{center}
\caption{
The energy spectrum   ($l=2$ mode) of the
gravitational waves for the case of the polytropic star
with a small compactness parameter. We set the
polytropic index $n = 0.5$  and the center of the
density 
$\rho_{\rm c} = 388.097 \times 10^{15} $  g / cm$^{3}$, which may
be implausible but is necessary to obtain an ultracompact
star. The solid and dotted lines correspond to  
$(E, L)=(2.38 \mu,12.0 \mu M)$ and $(1.01 \mu,4.5 \mu M)$,
respectively.
This figure shows the same features as  Fig. 5.
}
\label{fig:polytropic0.5}
\end{figure}

Next, we discuss the second type stellar model, i.e., an ultracompact star. 
To construct an ultracompact star with 
 a polytropic equation of 
state, 
we set 
the polytropic index $n = 0.5$, 
the constant $K = 100 \ {\rm km^4}$, and choose
 the  central energy density 
$\rho_{\rm c} = 388.097 \times 10^{15} \ {\rm g / cm^{3}}$,  
for which the compactness  parameter $R / M$ turns out to be  the
smallest.  For such a choice, the radius, mass, and compactness
parameters of the star are 
$R = 1.136 \ {\rm km}$, 
$M = 0.296 M_{\odot}$, and 
$R  =  2.597 M$, respectively.
Although this seems very implausible for a realistic star,
we will use this solution in our analysis to study the dependence of
the equation of state on the energy spectrum.
 In Fig. \ref{fig:polytropic0.5}, we show the
energy spectra  for two  trajectories of a test
particle. One is the case that a particle goes into the  inside of
the potential barrier,  while the other is the case that a particle
passes through only  outside the potential barrier.
We find the same feature as that of the  uniform density star with  
$R \lesssim 3 M$.
The difference is found in the frequencies of quasinormal modes
because of the different shapes of the potential.
Then, we find many small and sharp peaks in the spectrum at the
positions different from those for a uniform density star.
Consequently, we may be able to distinguish the inside structure of
an ultracompact star by observing the energy spectrum of
emitted gravitational waves.

%%%%%%%%%%%%%%%%%%%%%%%%%%%%%%%%%%%%%%%%%%%%%%%%%
\section{Concluding Remarks}
\label{sec:Discussion}
%%%%%%%%%%%%%%%%%%%%%%%%%%%%%%%%%%%%%%%%%%%%%%%%%

We have studied the axial modes of gravitational waves 
from a test particle scattered by a spherically symmetric relativistic
 star. We have considered both a uniform density star and a polytropic
star. We find that the energy spectrum 
depends mainly  on  the  shape  of the effective potential
for gravitational waves, i.e.,
whether its minimum exists or not.

When there is no minimum  (e.g., a regular type of neutron star with a
compactness parameter 
$R/M \gtrsim 3$),
the energy spectrum shows only a single global peak, 
which corresponds to the orbital frequency of a test particle at the
turning point.
Gravitational waves are mostly 
emitted  around the turning point, where 
the effect of the gravitational field is strongest.

While if  a minimum of the effective potential exists 
(i.e., some exotic relativistic star with small compactness parameter 
$R/M \lesssim 3$),
the energy spectrum shows a variety in its structure in addition to 
the global peak as the same as that found in the previous case.
We find small and sharp periodic peaks 
in the region lower  than 
$\omega_{\rm max}$ (the maximum value of the effective
potential), which correspond to an excitation of the axial quasinormal
modes. The existence of the potential minimum allows
the existence of a quasibound state of gravitational waves, i.e.,
quasinormal modes.
Then those modes are excited by a scattered particle.
This becomes more conspicuous when the particle gets into the inside
region beyond the potential barrier.
We also find small periodic peaks in the frequency region beyond
$\omega_{\rm max}$, which may come from an interference 
 between  two waves reflected by the potential barrier and the
infinite wall near the center of the star. 

From the observation of the energy spectrum of the axial mode, what we can 
determine? Naively speaking, we may not find any information about
the constituent of the star, because the axial mode does not couple to
matter fluid.  This is true for the first type stellar model, i.e., a
regular neutron star. Emitted gravitational waves are mostly
determined by the particle orbit.
 The case of black holes also belongs to this
case, although there is a small difference in the high frequency part
because of the difference of the boundary condition.
However, if a star is compact enough to make a potential minimum ($R
\lesssim 3M$), emitted gravitational waves show much more abundant
information. It may determine a compactness parameter $R/M$. We also
find
quasinormal mode frequencies, which depend on the shape of the
potential, and then could depend on the equation of state.

In the axial mode case, however, in general,
we cannot have direct information about the matter fluid, in particular
about  the  equation of state.
The polar mode of gravitational waves will provide us a more conspicuous
dependence of matter property. Work on the polar modes is under
way.

%%%%%%%%%%%%%%%%%%%%%%%%
\acknowledgments
%%%%%%%%%%%%%%%%%%%%%%%%
We would like to thank Yasufumi Kojima for 
kind suggestions at the beginning of this study.  
The Numerical Computations were mainly
performed by 
the FUJITSU-VX vector computer at the Media Network
Center, Waseda University. This work 
was supported partially by a JSPS Grant-in-Aid (No.
5689 and Specially Promoted Research No. 08102010), and by  the
Waseda University Grant  for Special Research Projects.

\newpage
\appendix
%%%%%%%%%%%%%%%%%%%%%%%%%%%%%%%%%%%%%%%%%%%%%%%%%
\section*{Numerical techniques solving the linearized Einstein equations}
%%%%%%%%%%%%%%%%%%%%%%%%%%%%%%%%%%%%%%%%%%%%%%%%%

Here, we present numerical techniques solving the wave equations.
 We improved  Borrelli's method\cite{Borrelli} to guarantee 
numerical accuracy.

For the interior region, 
the numerical error of the wave function accumulates and becomes
maximal  at the surface,  because  the pressure vanishes there. In order
to overcome this difficulty,  we solve the wave equation both  from the
center and  from the surface, and
connect two wave functions 
at some inner point, e.g., $r=R/2$.
For the exterior solution, 
we integrate the wave equation 
from the surface to infinity, by which we can reduce the numerical error.
In the present model, the test particle mainly  emits gravitational
waves at the turning point, where we have to solve the wave equation very
carefully because an apparent divergence appears.
We shall discuss these three methods in order.

%%%%%%%%%%%%%%%%%%%%%%%%
\subsection{Interior region of the star}
\label{subsec:app_interior}
%%%%%%%%%%%%%%%%%%%%%%%%

Here we show how to solve the 
interior wave equation (\ref{eq:inside_perturbation_equation}).
Transforming  the wave function $X_{l\omega}^{\rm (int)}$ to
$
Z_{l\omega} = r X_{l\omega}^{\rm (int)}, 
%\label{eq:transZX}
$
%%%%%%%%%
we find the  perturbation equation for $Z_{l\omega}$ 
as
%%%%%%%%%
\begin{eqnarray}
\frac{d^2 Z_{l\omega}}{dr^2} 
- \frac{e^{\lambda}}{r} 
\left[ 2 - \frac{6M(r)}{r} - 4\pi r^2 (P - \rho) \right]
\frac{dZ_{l\omega}}{dr}
+ e^{\lambda - \nu} \left( \omega^2 -\frac{2n}{r^2}e^{\nu}  \right) 
Z_{l\omega} = 0.
\label{eq:near_center_Z_equation}
\end{eqnarray}
%%%%%%%%%
From the regularity condition at the center, 
we obtain the asymptotic behavior at $r \sim 0$ as
%%%%%%%%%
\begin{eqnarray}
Z_{l\omega} = \eta_{l\omega} 
\left\{ r^{l+2} + \frac{1}{2(2l+3)} \left[ 4 \pi (l + 2)
\left( \frac{1}{3} (2l - 1) \rho_{\rm c} - P_{\rm c} \right)
- \omega^2 e^{- \nu_{\rm c}} \right] r^{l + 4} + O(r^{l+6}) \right\},
\label{eq:near_center_Z}
\end{eqnarray}
%%%%%%%%%
where  $\rho_{\rm c}$, $P_{\rm c}$, and $\nu_{\rm c}$ are 
density, pressure, and metric functions evaluated at the center,
respectively. In order to reduce the numerical error 
 at the surface,
we adopt the following procedure.

First, we construct the wave function $Z_{l\omega}^{(0)}$
  by integrating Eq. (\ref{eq:near_center_Z_equation}) from the center with
the asymptotic behavior,  Eq. (\ref{eq:near_center_Z}),
to a middle point ($r = R/2$).

Then, we find two independent wave functions 
$Z^{(1)}_{l\omega}, Z^{(2)}_{l\omega}$  
by integrating
Eq. (\ref{eq:near_center_Z_equation}) from the surface
to the middle point.
The boundary conditions of those wave functions at the surface are
given as   $(Z^{(1)}_{l\omega}, dZ^{(1)}_{l\omega}/dr) = (1, 0)$, 
$(Z^{(2)}_{l\omega}, dZ^{(2)}_{l\omega}/dr) = (0, 1)$.
By those two independent solutions, 
we construct the wave function $Z_{l\omega}(r)$ as
%%%%%%%%%
\begin{eqnarray}
Z_{l\omega}(r) = 
a_{l\omega} Z_{l\omega}^{(1)} (r) + b_{l\omega} Z_{l\omega}^{(2)} (r),
\label{eq:SolZ}
\end{eqnarray}
%%%%%%%%%
where the coefficients  $a_{l\omega}$ and  $b_{l\omega}$ are fixed
by junction conditions at $r=R/2$.
Since the wave functions constructed above must
be continuous and smooth  at
$r=R/2$, we find the junction conditions 
%%%%%%%%
\begin{eqnarray}
Z_{l\omega}^{(0)} \left( \frac{R}{2} \right) &=& 
a_{l\omega} Z_{l\omega}^{(1)} \left( \frac{R}{2} \right) + 
b_{l\omega} Z_{l\omega}^{(2)} \left( \frac{R}{2} \right), 
\\
 \frac{dZ_{l\omega}^{(0)}}{dr} \left( \frac{R}{2} \right) &=&
a_{l\omega}  \frac{dZ_{l\omega}^{(1)}}{dr} \left( \frac{R}{2} \right)
+ b_{l\omega}   \frac{dZ_{l\omega}^{(2)}}{dr} \left( \frac{R}{2}
\right).
\end{eqnarray}
%%%%%%%%
From these two conditions, we find $a_{l\omega}$ and $b_{l\omega}$, and
then 
 the interior wave function $Z_{l\omega}$.

We finally find values of $X_{l\omega}^{\rm (int)}$ and 
$dX_{l\omega}^{\rm (int)}/dr^{*}$ at the surface $r=R$ as 
\begin{eqnarray}
%%%%%%%%
X_{l\omega}^{\rm (int)}(R^{*}) &=& \frac{1}{R} Z_{l\omega}(R), 
\label{eq:initial_value_at_surface_X} \\
 \frac{d X_{l\omega}^{\rm (int)}}{dr^{*}} \left(R^{*}\right) 
&=& \left( 1 - \frac{2M}{R} \right) 
\left( \frac{1}{R}  \frac{dZ_{l\omega}}{dr} \left( R \right) - 
\frac{1}{R^2} Z_{l\omega}(R) \right),
\label{eq:initial_value_at_surface_dXdtr}
\end{eqnarray}
%%%%%%%%
where $R^{*}$ is the tortoise coordinate of the surface $R$.

%%%%%%%%%%%%%%%%%%%%%%%%
\subsection{Exterior region of the star}
\label{subsec:app_exterior}
%%%%%%%%%%%%%%%%%%%%%%%%

Next we construct the exterior wave function.
In order to use the Green's function method,
we first consider the homogeneous wave function 
$X_{lm\omega}^{\rm (ext)(0)}$, which satisfies the wave equation 
%%%%%%%%%
\begin{equation}
\frac{d^2 X_{lm\omega}^{\rm (ext)(0)}}{d {r^{*}}^2} 
+ \left( \omega^2 - V_{l}^{\rm (ext)} \right)
X_{lm\omega}^{\rm (ext)(0)} = 0, 
\label{eq:outside_perturbation_eq_homogeneout}
\end{equation}
%%%%%%%%%
and the boundary condition at the surface, which we have already
evaluated  in the previous subsection, that is,
%%%%%%%%%
\begin{eqnarray}
X_{lm\omega}^{\rm (ext)(0)}(R^{*}) &=& 
X_{l\omega}^{\rm (int)}(R^{*}), 
\label{boundary1}
\\
\displaystyle 
\frac{d X_{lm\omega}^{\rm (ext)(0)}}{dr^{*}} \left(R^{*}\right) 
 &=& 
\displaystyle 
\frac{d X_{l\omega}^{\rm (int)}}{d r^{*}} \left(R^{*}\right).
\label{boundary2}
\end{eqnarray}
%%%%%%%%%

Then, we construct a particular solution
$X_{lm\omega}^{\rm (ext)(1)}$ with a source term, 
which equation
is described as
%%%%%%%%%
\begin{equation}
\frac{d^2 X_{lm\omega}^{\rm (ext)(1)}}{d {r^{*}}^2} 
+ \left( \omega^2 - V_{l}^{\rm (ext)} \right)
X_{lm\omega}^{\rm (ext)(1)} = S_{lm\omega}^{\rm (ext)},
\label{eq:outside_perturbation_eq_inhomogeneout}
\end{equation}
%%%%%%%%%
where $S_{lm\omega}^{\rm (ext)}$ comes from the motion of a test
particle.
At the surface, the solution $X_{lm\omega}^{\rm (ext)(1)}$ satisfies the
following condition: 
%%%%%%%%%
\begin{eqnarray}
X_{lm\omega}^{\rm (ext)(1)}(R^{*}) &=& 0, \\
\displaystyle 
\frac{d X_{lm\omega}^{\rm (ext)(1)}}{dr^{*}} 
\left(R^{*}\right) &=& 0.
\end{eqnarray}
%%%%%%%%%

For the homogeneous equation 
(\ref{eq:outside_perturbation_eq_homogeneout}), 
we have two independent solutions 
 $u_{lm\omega}^{\rm (out)}(r^{*})$ and
$u_{lm\omega}^{\rm (in)}(r^{*})$, 
which correspond to 
outgoing and incoming waves at infinity, respectively.
Then, we describe the wave function $X_{lm\omega}^{\rm (ext)(0)}$ 
by these two independent solutions as 
%%%%%%%%%
\begin{eqnarray}
X_{lm\omega}^{\rm (ext)(0)}(r^{*}) = 
\alpha_{lm\omega} u_{lm\omega}^{\rm (out)}(r^{*}) + 
\beta_{lm\omega} u_{lm\omega}^{\rm (in)}(r^{*}),
\end{eqnarray}
%%%%%%%%%
where $\alpha_{lm\omega}$ and $\beta_{lm\omega}$ 
are complex constants, which are fixed by
Eqs. (\ref{boundary1}), (\ref{boundary2}).

The Green's function $G(r^{*}, s^{*})$ 
for Eq. (\ref{eq:outside_perturbation_eq_inhomogeneout})
is also constructed by $u_{lm\omega}^{\rm (out)}$ and 
$u_{lm\omega}^{\rm (in)}$ as 
%%%%%%%%%
\begin{eqnarray}
G(r^{*}, s^{*}) = \frac{1}{W} \Bigl[
- u^{\rm (out)}(r^{*}) u^{\rm (in)}(s^{*})
+ u^{\rm (in)}(r^{*}) u^{\rm (out)}(s^{*}) 
\Bigr] \theta(r^{*} - s^{*}) \qquad (r^{*} > s^{*}),
\end{eqnarray}
%%%%%%%%%
where $W$ is the Wronskian and $\theta(x)$ is the Heaviside function.
The wave function  $X_{lm\omega}^{\rm
(ext)(1)}$ is obtained by integration as
%%%%%%%%%
\begin{equation}
X_{lm\omega}^{\rm (ext)(1)}(r^{*}) = 
\int_{R^{*}}^{\infty} G(r^{*}, s^{*}) 
S_{lm\omega}^{\rm (ext)} (s^{*}) ds^{*}.
\end{equation}
%%%%%%%%%

Then, the solution for the exterior wave 
equation  (\ref{eq:outside_perturbation_equation})
with the same boundary conditions as Eqs. (\ref{boundary1}), (\ref{boundary2})
is given as 
%%%%%%%%%
\begin{eqnarray}
X_{lm\omega}^{\rm (ext)}(r^{*}) = 
\alpha_{lm\omega} u_{\rm (out)}(r^{*})
+ \beta_{lm\omega} u_{\rm (in)}(r^{*})
+ \int_{R^{*}}^{\infty} G(r^{*}, s^{*}) 
S_{lm\omega}^{\rm (ext)} (s^{*}) ds^{*}
\label{eq:outside_perturbation_equation_general_solution}.
\end{eqnarray}
%%%%%%%%%

The asymptotic behavior of  $X_{lm\omega}^{\rm (ext)}(r^{*})$ as $r
\rightarrow \infty$ is given by 
%%%%%%%%%
\begin{eqnarray}
X_{lm\omega}^{\rm (ext)}(r^{*}) \to (\alpha_{lm\omega} +
\rho_{lm\omega})  e^{i
\omega r^{*}}  + (\beta_{lm\omega} + \sigma_{lm\omega}) e^{-i \omega
r^{*}}
\qquad (r^{*} \to \infty),
\end{eqnarray}
%%%%%%%%%
where 
$\rho_{lm\omega}$ and $\sigma_{lm\omega}$ are defined by
%%%%%%%%%
\begin{eqnarray}
\rho_{lm\omega} &=& 
- \frac{1}{W} \int_{R^{*}}^{\infty} 
u_{lm\omega}^{\rm (in)} (s^{*}) 
S_{lm\omega}^{\rm (ext)} (s^{*}) ds^{*}, \\
\sigma_{lm\omega} &=& 
\frac{1}{W} \int_{R^{*}}^{\infty} 
u_{lm\omega}^{\rm (out)} (s^{*}) 
S_{lm\omega}^{\rm (ext)} (s^{*}) ds^{*}.
\end{eqnarray}
%%%%%%%%%
Since there is no incoming wave, we find the boundary condition such
that 
%%%%%%%%%
\begin{eqnarray}
\beta_{lm\omega} + \sigma_{lm\omega} = 0.
\label{eq:boundary_condition_infinity}
\end{eqnarray}
%%%%%%%%%
Then, the amplitude $A_{lm\omega}$ of gravitational waves 
at infinity is given by 
%%%%%%%%%
\begin{eqnarray}
A_{lm\omega} = \alpha_{lm\omega} + \rho_{lm\omega}.
\end{eqnarray}
%%%%%%%%%

In order to extract the amplitude $A_{lm\omega}$ from our solution, 
we use the relation
%%%%%%%%%
\begin{eqnarray}
- \beta_{lm\omega} \rho_{lm\omega} 
+ \alpha_{lm\omega} \sigma_{lm\omega} &=&
\frac{1}{W} \int_{R^{*}}^{\infty} 
\Bigl[ \alpha_{lm\omega} u_{lm\omega}^{\rm (out)} (s^{*})
+ \beta_{lm\omega} u_{lm\omega}^{\rm (in)} (s^{*}) \Bigr]
S_{lm\omega}^{\rm (ext)} (s^{*}) ds^{*} 
\nonumber \\
&=& \frac{1}{W} \int_{R^{*}}^{\infty}
X_{lm\omega}^{\rm (ext)(0)}(s^{*}) 
S_{lm\omega}^{\rm (ext)} (s^{*}) ds^{*}.
\label{eq:boundary_condition_infinity2}
\end{eqnarray}
%%%%%%%%%
With Eqs. (\ref{eq:boundary_condition_infinity}) and
(\ref{eq:boundary_condition_infinity2}), 
we finally obtain the amplitude  as
%%%%%%%%%
\begin{eqnarray}
A_{lm\omega} = 
\alpha_{lm\omega} + \rho_{lm\omega} = 
- \frac{1}{\beta_{lm\omega}} \frac{1}{W}
\int_{R^{*}}^{\infty} 
X_{lm\omega}^{\rm (ext)(0)}(s^{*}) 
S_{lm\omega}^{\rm (ext)} (s^{*}) ds^{*},
\label{eq:amplitude_result}
\end{eqnarray}
%%%%%%%%%
where $\beta_{lm\omega}$ is already obtained up to an unknown constant
$\eta_{l \omega}$, which  is canceled out with that of
$X_{lm\omega}^{\rm (ext)(0)}$ in Eq. (\ref{eq:amplitude_result})

%%%%%%%%%%%%%%%%%%%%%%%%
\subsection{Numerical method to integrate near the turning point of a test particle orbit}
\label{subsec:app_turning_point}
%%%%%%%%%%%%%%%%%%%%%%%%
When a test particle is scattered by a  star, 
the orbit includes a turning point ($\hat{R}_{\rm min}$ or 
 $\hat{R}_{\rm min}^{*}$ by the tortoise coordinate), 
where the  radial velocity $d \hat{R}/d \tau $ vanishes.
The source term $S_{lm\omega}^{\rm (ext)}$ in Eq. (\ref{eq:amplitude_result})
is given by Eq. (\ref{source}), where $Q_{lm}(\omega,r)$ and
$D_{lm}(\omega,r) $, defined by Eqs. (\ref{eq:def_Q}) and (\ref{eq:def_D}), 
for the present orbit
($\hat{\Theta} = \pi /2$) are described as 
%%%%%%%%%
\begin{eqnarray}
Q_{lm}(\omega,r) &=& 
- C_{lm}\sqrt{\frac{2}{l(l+1)}} \frac{\mu}{r \left( 1 - 2M/r \right)} 
\int_{-\infty}^{\infty} d \tau
\frac{d\hat{R}}{d \tau} \delta {\bf (} r - \hat{R}(\tau) {\bf )} \nonumber \\
& & \qquad \qquad \qquad \qquad \qquad \qquad \times 
\frac{\tilde{L}}{\hat{R}^2}
\frac{d P_{lm}}{d \hat{\Theta}}  \left({\pi\over 2}\right)  
e^{i(\omega \hat{T} - m \hat{\Phi})}, \\
D_{lm}(\omega,r) &=& 
\frac{-imC_{lm}}{\sqrt{2n(n+1)}} \mu \int_{-\infty}^{\infty} 
d \tau \delta (r - \hat{R}(\tau)) 
\frac{\tilde{L}^2}{\hat{R}^4} \frac{d P_{lm}}{d \hat{\Theta}} 
 \left({\pi\over 2}\right)  e^{i(\omega \hat{T} - m \hat{\Phi})},
\label{D_lm}
\end{eqnarray}
%%%%%%%%%
where 
%%%%%%%%%
\begin{equation}
C_{lm} = (-1)^{(m + |m|)/2} 
\sqrt{\frac{(2l + 1)}{4 \pi} \frac{(l - |m|)!}{(l + |m|)!}},
\end{equation}
%%%%%%%%%
which is a normalization constant 
of a spherical harmonics $Y_{lm}$,
and $P_{lm}(\cos \theta)$ is the associated Legendre function.

Integration  over the proper time $\tau$ gives the term of $(d
\hat{R}/d \tau )^{-1}$ in Eq. (\ref{D_lm}), which diverges at the turning
point, and then 
the integration
%%%%%%%%%
\begin{eqnarray}
\frac{1}{W} \int_{R^{*}}^{\infty}
X_{lm\omega}^{\rm (ext)(0)} (s^{*}) 
S_{lm\omega}^{\rm (ext)} (s^{*}) ds^{*}
= \frac{1}{W} \int_{\hat{R}_{\rm min}^{*}}^{\infty}
X_{lm\omega}^{\rm (ext)(0)} (s^{*}) 
S_{lm\omega}^{\rm (ext)} (s^{*}) ds^{*}
\label{integration}
\end{eqnarray}
%%%%%%%%%
diverges apparently. 
However, this is not a physical divergence, but rather 
comes from the change
of integration variables 
from the proper time $\tau$ to the radius $r$.
This transformation  becomes singular at the turning point because
$d\hat{R} /d\tau =0$ . In order to avoid this difficulty, we shall use a
time coordinate $\hat{T}$ in the integration.

Since the trajectory  before and after the turning point is symmetric,
we choose  the initial position of the particle $\hat{T} = 0$,
$\hat{\Phi} = 0$ at  the turning point and integrate from there to
infinity.  
The integral, Eq. (\ref{integration}), is then written as 
%%%%%%%%%
\begin{eqnarray}
&& \frac{1}{W} \int_{\hat{R}_{\rm min}}^{\infty} 
ds^{*} X_{lm\omega}^{\rm (ext)(0)} 
S_{lm\omega}^{\rm (ext)} (s^{*}) = 
\frac{1}{W} \int_{0}^{\infty} d \hat{T} 
\Biggl[ X_{lm\omega}^{\rm (ext)(0)} (\hat{T}) 
\left[ S_{lm}^{(0)}(\hat{T}) + 
S_{lm}^{(1)}(\hat{T}) \right] \nonumber 
\\
&& \qquad \qquad \qquad \qquad \qquad \qquad \qquad 
+ \hat{R} \left( 1 - \frac{2M}{\hat{R}} \right)
\frac{1}{d \hat{R}/ d \hat{T}}  \frac{d X_{lm\omega}^{\rm
(ext)(0)}}{d \hat{T}} S_{lm}^{(1)} (\hat{T})
\Biggr], 
\label{eq:source_term_integral_t} \\
&& \qquad \hat{S}_{lm}^{(0)} = \frac{8 \pi}{\tilde{E}} \Biggl[
\sqrt{\frac{2}{l(l + 1)}} \left( 1- \frac{2M}{\hat{R}} \right)^2 
\hat{Q}_{lm} \nonumber 
\\
&& \qquad \qquad \qquad \qquad \qquad 
+ \sqrt{\frac{2}{l(l + 1)(l - 1)(l + 2)}} \frac{2M}{\hat{R}}
\left( 1 - \frac{2M}{\hat{R}} \right) \hat{D}_{lm}
\Biggr], 
\\
&& \qquad \hat{S}_{lm}^{(1)} = -\frac{8 \pi}{\tilde{E}} 
\sqrt{\frac{2}{l(l + 1)(l - 1)(l + 2)}} 
\left( 1 - \frac{2M}{\hat{R}} \right) \hat{D}_{lm}, 
\\
&& \qquad \hat{Q}_{lm} = - 2 i C_{lm} \sqrt{\frac{2}{l(l + 1)}} 
\frac{\mu \tilde{L}}{\hat{R}^3 (1 - 2M/\hat{R})} \gamma^{1/2}
 \frac{d P_{lm}}{d \hat{\Theta}} \left({\pi\over 2}\right) 
\sin (\omega \hat{T} - m \hat{\Phi}), 
\\
&& \qquad \hat{D}_{lm} =  
\frac{- 4imC_{lm}}{\sqrt{2l(l + 1)(l - 1)(l + 2)}}
\frac{\mu \tilde{L}^2}{\hat{R}^4} 
 \frac{d P_{lm}}{d \hat{\Theta}}\left({\pi\over 2}\right) 
\cos (\omega \hat{T} - m \hat{\Phi}),
\end{eqnarray}
%%%%%%%%%
where $X_{lm\omega}^{\rm
(ext)(0)}(\hat{T}) = X_{lm\omega}^{\rm
(ext)(0)} {\bf (} \hat{R}(\hat{T}) {\bf )}$ and $d X_{lm\omega}^{\rm
(ext)(0)}/d \hat{T}$ is its time derivative.
However, the term  in Eq. 
(\ref{eq:source_term_integral_t}),
%%%%%%%%%
\begin{eqnarray}
\frac{1}{d \hat{R} / d \hat{T}}  \frac{d X_{lm\omega}^{\rm (ext)(0)}}{d
\hat{T}} = \frac{d X_{lm\omega}^{\rm (ext)(0)}}{d\hat{R}},
\end{eqnarray}
%%%%%%%%%
is still apparently divergent because 
$d \hat{R}/d \hat{T}$ vanishes at the turning point.

To avoid this difficulty, we 
expand $X_{lm\omega}^{\rm (ext)(0)}$, 
$dX_{lm\omega}^{\rm (ext)(0)}/d\hat{R}$
near the turning point.
From the equation of motion, we find
%%%%%%%%%
\begin{eqnarray}
\frac{d \hat{R}}{d \hat{T}} \propto (\hat{R}- \hat{R}_{\rm min})^{1/2} +
O(\hat{R} - \hat{R}_{\rm min}),
\end{eqnarray}
%%%%%%%%%
which is integrated as 
%%%%%%%%%
\begin{eqnarray}
\hat{R} - \hat{R}_{\rm min} = a_{0} \hat{T}^{2} + O(\hat{T}^{4}),
\end{eqnarray}
%%%%%%%%%
where $a_{0}$ is an integration constant.
Then, we find the wave function 
 near the turning point as
%%%%%%%%%
\begin{eqnarray}
X_{lm\omega}^{\rm (ext)(0)} &=&  X_{lm\omega}^{\rm (ext)(0)}(\hat{R}_{\rm min})  + 
 \frac{d X_{lm\omega}^{\rm (ext)(0)}}{dr} \left(\hat{R}_{\rm min}\right)   (\hat{R}
-\hat{R}_{\rm min}) \nonumber \\ &=& X_{lm\omega}^{\rm (ext)(0)}(\hat{R}_{\rm min})  + 
\frac{a_{0}}{1 - 2M/\hat{R}_{\rm min}} 
\frac{d X_{lm\omega}^{\rm (ext)(0)}}{dr^{*}} 
\left(\hat{R}_{\rm min}^{*}\right) \hat{T}^{2}, \\
\frac{dX_{lm\omega}^{\rm (ext)(0)}}{d\hat{R}} &=& 
 \frac{d X_{lm\omega}^{\rm (ext)(0)}}{dr}  \left(\hat{R}_{\rm min}\right)   +
 \frac{d^2 X_{lm\omega}^{\rm (ext)(0)}}{d r^2}
\left(\hat{R}_{\rm min}\right)   (\hat{R} - \hat{R}_{\rm min}) \nonumber \\
&=& \frac{1}{1 - 2M/\hat{R}_{\rm min}}   \frac{d X_{lm\omega}^{\rm
(ext)(0)}}{dr^{*}} 
\left(\hat{R}_{\rm min}^{*}\right) \nonumber \\
& &    +  \frac{a_{0}}{(1 - 2M/\hat{R}_{\rm min})^2} 
\left( - \frac{2M}{\hat{R}_{\rm min}^2}    \frac{d X_{lm\omega}^{\rm
(ext)(0)}}{dr^{*}} 
\left(\hat{R}_{\rm min}^{*}\right)
+  \frac{d^2 X_{lm\omega}^{\rm (ext)(0)}}{d {r^{*}}^2} 
\left(\hat{R}_{\rm min}^{*}\right) \right) \hat{T}^2,
\end{eqnarray}
%%%%%%%%%
where $X_{lm\omega}^{\rm (ext)(0)}$, 
$d X_{lm\omega}^{\rm (ext)(0)}/dr^{*}$, and
$d^2 X_{lm\omega}^{\rm (ext)(0)}/d {r^{*}}^2$ 
at the turning point $\hat{R}_{\rm min}^{*}$ are found easily 
by integration of Eq. 
(\ref{eq:outside_perturbation_eq_homogeneout}).
Since those functions are regular at the turning point, we can evaluate
Eq. (\ref{eq:source_term_integral_t}) without any difficulty.

%%%%%%%%%%%%%%%%%%%%%%%%%%%%%%%%%%%%%%%%%%%%%%%%%
%%%
%%%   Reference
%%%
%%%%%%%%%%%%%%%%%%%%%%%%%%%%%%%%%%%%%%%%%%%%%%%%%

\newpage
%%%%%%%%%%%%%%%%%%%%%%%%%%%%%%%%%%%%%%%%%%%%%%%%%
%%%
%%%   Figure Captions
%%%
%%%%%%%%%%%%%%%%%%%%%%%%%%%%%%%%%%%%%%%%%%%%%%%%%

\newpage
%%%%%%%%%%%%%%%%%%%%%%%%%%%%%%%%%%%%%%%%%%%%%%%%%
%%%
%%%   Table Captions
%%%
%%%%%%%%%%%%%%%%%%%%%%%%%%%%%%%%%%%%%%%%%%%%%%%%%
\begin{table}
\caption{
Axial ``quasinormal" modes
for a  uniform density star with 
$R   = 2.26 M$.
We use the continued fraction expansion method, which was
first used to calculate black hole quasinormal modes by Leaver [24]
%\cite{Leaver}
, and 
$w$ modes of a polytropic star by Leins, Nollert, and Soffel [16]
%\cite{LNS}
.
Our results agree quite well with those of  
Kokkotas, who adopted a different method [20].
%\cite{Kokkotas}
Our imaginary part has opposite sign to Kokkotas's result, 
because our definition of the Fourier transformation is opposite to his.
}
\label{tab:QNM_Urm2.26}
\begin{center}

\begin{tabular}{c c  c c}
\multicolumn{2}{c}{Our results} &
\multicolumn{2}{c}{Kokkotas's results}\\ 
${\rm Re}[M \omega]$ & $-{\rm Im}[M \omega]$ &
${\rm Re}[M \omega]$ & ${\rm Im}[M \omega]$ \\ \hline
$0.1090$ & $1.2399 \times 10^{-9}$ & $0.1091$ & $1.2388 \times 10^{-9}$ \\
$0.1484$ & $3.9495 \times 10^{-8}$ & $0.1484$ & $3.9494 \times 10^{-8}$ \\
$0.1876$ & $5.4678 \times 10^{-7}$ & $0.1876$ & $5.4701 \times 10^{-7}$ \\
$0.2267$ & $4.8526 \times 10^{-6}$ & $0.2267$ & $4.8528 \times 10^{-6}$ \\
$0.2654$ & $3.2318 \times 10^{-5}$ & $0.2654$ & $3.2316 \times 10^{-5}$ \\
$0.3036$ & $1.7234 \times 10^{-4}$ & $0.3036$ & $1.7236 \times 10^{-4}$ \\
$0.3410$ & $7.3017 \times 10^{-4}$ & $0.3411$ & $7.3003 \times 10^{-4}$ \\
$0.3777$ & $2.2982 \times 10^{-3}$ & $0.3777$ & $2.2982 \times 10^{-3}$ \\
$0.4144$ & $5.1672 \times 10^{-3}$ & $0.4144$ & $5.1490 \times 10^{-3}$ \\
$0.4516$ & $8.7987 \times 10^{-3}$ & $0.4516$ & $8.7685 \times 10^{-3}$ \\
$0.4896$ & $1.2480 \times 10^{-2}$ & $0.4896$ & $1.2490 \times 10^{-2}$ \\
$0.5282$ & $1.5818 \times 10^{-2}$ & $0.5282$ & $1.5804 \times 10^{-2}$ \\
$0.5674$ & $1.8702 \times 10^{-2}$ & $0.5675$ & $1.8710 \times 10^{-2}$ \\
$0.6074$ & $2.1187 \times 10^{-2}$ & $0.6074$ & $2.1208 \times 10^{-2}$ \\
$0.6481$ & $2.3368 \times 10^{-2}$ & $0.6481$ & $2.3349 \times 10^{-2}$ \\
\end{tabular}
\end{center}
\end{table}
\end{document}